\documentclass[nonacm,screen]{acmart}
\usepackage[ruled,vlined,linesnumbered]{algorithm2e}
\SetFuncSty{textsc}
\SetCommentSty{itshape}
\SetKwComment{Comment}{$\triangleright$\ }{}
\SetKwFunction{GetSubgraph}{GetSubgraph}
\SetKwProg{Fn}{Function}{:}{}

\usepackage{caption}
\usepackage{subcaption}
\usepackage{multirow}
\usepackage{booktabs}

\usepackage{tikz}
\usepackage{enumitem} 
\definecolor{whitesmoke}{rgb}{0.96, 0.96, 0.96}
\newcommand*\circleda[1]{\tikz[baseline=(char.base)]{%
            \node[shape=circle,fill=whitesmoke,draw,inner sep=1pt, minimum size=3mm] (char) {#1};}}

\usepackage{booktabs}
\usepackage{makecell}

\renewcommand{\figurename}{Fig.}
\renewcommand{\sectionname}{Section}
\newcommand{\p}[1]{\phantom{#1}}

\AtBeginDocument{%
  }

\copyrightyear{2026}
\acmYear{2026}
\acmDOI{}

\begin{document}

\title[ParaQAOA for Handling Large-Scale Max-Cut Problems]{ParaQAOA: Efficient Parallel Divide-and-Conquer QAOA for Large-Scale Max-Cut Problems Beyond 10,000 Vertices}
\author{Po-Hsuan Huang}
\email{aben20807@gmail.com}
\orcid{0000-0002-7458-9634}
\affiliation{
  \department {Department of Computer Science and Information Engineering}
  \institution{National Taiwan University}
  \city{Taipei}
  \country{Taiwan}
}
\author{Xie-Ru Li}
\email{p76134587@gs.ncku.edu.tw}\
\orcid{0009-0000-5354-5248}
\affiliation{%
  \department {Department of Computer Science and Information Engineering}
  \institution{National Cheng Kung University}
  \city{Tainan}
  \country{Taiwan}
}

\author{Chi Chuang}
\email{ab321013@gmail.com}\
\orcid{0009-0003-2956-8222}
\affiliation{%
  \department {Department of Computer Science and Information Engineering}
  \institution{National Cheng Kung University}
  \city{Tainan}
  \country{Taiwan}
}

\author{Chia-Heng Tu}
\authornote{Corresponding author}
\email{chiaheng@ncku.edu.tw}
\orcid{0000-0001-8967-1385}
\affiliation{%
  \department {Department of Computer Science and Information Engineering}
  \institution{National Cheng Kung University}
  \city{Tainan}
  \country{Taiwan}
}

\author{Shih-Hao Hung}
\email{hungsh@csie.ntu.edu.tw}
\orcid{0000-0003-2043-2663}
\affiliation{
  \department {Department of Computer Science and Information Engineering}
  \institution{National Taiwan University}
  \city{Taipei}
  \country{Taiwan}
}

\renewcommand{\shortauthors}{P.-H. Huang et al.}

\begin{abstract}
Quantum Approximate Optimization Algorithm (QAOA) has emerged as a promising solution for combinatorial optimization problems using a hybrid quantum-classical framework. Among combinatorial optimization problems, the Maximum Cut (Max-Cut) problem is particularly important due to its broad applicability in various domains. While QAOA-based Max-Cut solvers have been developed, they primarily favor solution accuracy over execution efficiency, which significantly limits their practicality for large-scale problems. To address the limitation, we propose ParaQAOA, a parallel divide-and-conquer QAOA framework that leverages parallel computing hardware to efficiently solve large Max-Cut problems. ParaQAOA significantly reduces runtime by partitioning large problems into subproblems and solving them in parallel while preserving solution quality. This design not only scales to graphs with tens of thousands of vertices but also provides tunable control over accuracy-efficiency trade-offs, making ParaQAOA adaptable to diverse performance requirements. Experimental results demonstrate that ParaQAOA achieves up to 1,600x speedup over state-of-the-art methods on Max-Cut problems with 400 vertices while maintaining solution accuracy within 2\% of the best-known solutions. Furthermore, ParaQAOA solves a 16,000-vertex instance in 19 minutes, compared to over 13.6 days required by the best-known approach. These findings establish ParaQAOA as a practical and scalable framework for large-scale Max-Cut problems under stringent time constraints.

\end{abstract}




\maketitle

\section{Introduction}

Quantum computing is an emerging computational paradigm that extends beyond the scalability limits of classical computers. By harnessing quantum mechanical phenomena, such as superposition, entanglement, and interference~\cite{Feynman1982, Preskill2018quantumcomputingin}, quantum computers can achieve exponential speedups~\cite{shor,365701, PhysRevX.13.041041, Li2025} for specific computational tasks, which makes them well-suited for a range of computationally intensive problems in fields, such as cryptography, machine learning, and combinatorial optimization. Driven by rapid advances in both theoretical algorithms and hardware development, quantum computing has attracted significant global attention and is increasingly recognized as a key technology for next-generation computational solutions~\cite{Arute2019, Wu2025QOPS}.

Hybrid quantum-classical algorithms, among various quantum computing paradigms, have demonstrated remarkable suitability for deployment on current quantum hardware platforms by leveraging the complementary strengths of quantum and classical computational resources. These hybrid approaches exploit practical quantum advantage while mitigating the inherent limitations of contemporary quantum hardware. The Quantum Approximate Optimization Algorithm \cite{QAOA} is a notable example, which provides a robust theoretical foundation and has demonstrated practical performance for solving Quadratic Unconstrained Binary Optimization (QUBO) \cite{QUBO} problems across various application domains. In particular, QAOA's quantum-classical optimization framework has proven particularly effective for addressing combinatorial optimization problems faced in real-world applications.

The Maximum Cut problem is a classic application domain for QAOA-based optimization, as it serves both as a fundamental benchmark in computational complexity and as a practical optimization problem with real-world impact. The Max-Cut problem seeks a partition of a graph's vertices into two disjoint sets that maximizes the number of edges crossing between the sets. As a representative NP-hard combinatorial problem, Max-Cut has direct applications in various domains, such as VLSI circuit design optimization~\cite{maxcut_app4_vlsi}, social network community detection~\cite{maxcut_app1,maxcut_app2}, and wireless network frequency assignment~\cite{maxcut_app5_Wireless}. The problem's broad applicability and inherent computational difficulty call for efficient, scalable algorithms that can handle large-scale instances while preserving solution quality.

Many QAOA-based solutions have been developed for Max-Cut problems. Traditional QAOA implementations are resource-intensive as problem size scales, where circuit depth and gate complexity grow polynomially with the number of qubits. These resource demands render QAOA implementations infeasible on current quantum and classical hardware, thereby limiting their practical deployment for real-world applications. To address this, recent divide-and-conquer approaches~\cite{dcqaoa,QiQ,CQ} apply problem decomposition techniques, such as graph clustering, hierarchical partitioning, and recursive subdivision, coupled with quantum circuit simulations on classical computers. Unfortunately, while these methods reduce the computational load of subproblems and prioritize preserving solution quality (measured by \emph{approximation ratio} as introduced in \sectionname~\ref{sec:max_cut}), they typically increase overall execution time, which constrains their applicability to large problems. For instance, in our experiments in \sectionname~\ref{sec:performance_comparison}, Coupling QAOA~\cite{CQ} attains approximately 99\% of the approximation ratio on a 30-vertex graph but requires about eight hours to produce the Max-Cut result. This evidence shows that balancing computational efficiency and solution quality remains a central challenge for scaling QAOA to practical problem sizes.

In this work, we propose \emph{ParaQAOA}, a parallel divide-and-conquer QAOA framework that leverages modern parallel computing hardware to efficiently tackle large-scale Max-Cut problems. The ParaQAOA framework incorporates four key components. First, the framework employs a linear-time graph partitioning algorithm that reduces decomposition complexity and enables efficient handling of large graphs. Second, a parallelized execution pipeline is used to handle subproblem solving and solution reconstruction by leveraging modern parallel architectures to reduce overall runtime. Third, the framework provides a systematic parameterized design to better control parallel execution and manage the trade-off between execution efficiency and solution quality. Fourth, ParaQAOA introduces a unified metric that jointly evaluates solution quality and execution efficiency, and the metric enables consistent comparisons and informed trade-offs across different Max-Cut solutions. The key features of the framework are further detailed in \sectionname~\ref{sec:motivation}. The contributions of this work are as follows.

\begin{enumerate}    
    \item A parallel divide-and-conquer QAOA framework, ParaQAOA, is proposed to efficiently solve large-scale Max-Cut problems with parallel computing architectures on classical computers.
    To the best of our knowledge, ParaQAOA is the pioneering work that demonstrates an efficient solution for solving Max-Cut problems with over 10,000 vertices on classical computers and enables control over the trade-off between solution quality and execution efficiency, which is important when users aim to find a solution quickly even if it requires a sacrifice in solution quality.
    \item An important consideration, the trade-off between solution quality and execution efficiency, is introduced as a critical design aspect of a Max-Cut solver. This is a critical consideration when evaluating different Max-Cut solutions, especially for large-scale problems. To this end, we propose the Performance Efficiency Index (PEI), a novel evaluation metric that integrates approximation quality and runtime efficiency. Furthermore, the parameterized design of ParaQAOA enables control over the trade-off. 

    \item A series of experiments has been conducted to demonstrate the effectiveness of ParaQAOA. The experiments show that ParaQAOA effectively manages the trade-off between solution quality and execution efficiency. For instance, ParaQAOA achieves up to 1,600$\times$ speedup over state-of-the-art methods while maintaining approximation ratios within 2\% of the best-known solutions on 400-vertex instances (as medium-size problems). Additionally, the best-known prior approach would require 13.6 days to obtain a result for a 16,000-vertex graph, while ParaQAOA generates the result within 19 minutes. These results showcase the applicability of ParaQAOA to large-scale Max-Cut problems, particularly in scenarios with stringent time constraints.
\end{enumerate}

The remainder of this article is organized as follows. \sectionname~\ref{sec:background} provides the background on QAOA for Max-Cut, introduces the existing QAOA-based solutions to Max-Cut problems, analyzes their limitations, and elaborates the motivation for our proposed framework. \sectionname~\ref{sec:method} details the overall system design and the key components of the ParaQAOA framework. Experimental evaluation and analysis are presented in \sectionname~\ref{sec:evaluation} to demonstrate the effectiveness of our approach. Finally, \sectionname~\ref{sec:conclusion} concludes this work and discusses potential directions for future work.

\section{Background and Motivation}\label{sec:background}

This section establishes the theoretical foundations and motivation for the proposed ParaQAOA framework, with a focus on the Max-Cut problem and its relevance in quantum optimization studies in the literature. In \sectionname~\ref{sec:max_cut}, we introduce the formal definition of the Max-Cut problem. \sectionname~\ref{sec:qaoa} outlines how QAOA can be applied to solve Max-Cut instances, and \sectionname~\ref{sec:quantum_circuit_simulation} discusses the quantum circuit simulation for QAOA.
We then review existing QAOA-based approaches in \sectionname~\ref{sec:dc_qaoa} for solving the Max-Cut problem. We also summarize their strategies and limitations.
Finally, in \sectionname~\ref{sec:motivation}, we present our observations that motivated the development of ParaQAOA. These observations are based on the limitations of existing approaches, as discussed in \sectionname~\ref{sec:dc_qaoa}. Moreover, we highlight the features of our framework for handling large-scale Max-Cut problems.

\subsection{Maximum Cut Problem}\label{sec:max_cut}

The Max-Cut problem~\cite{Karp1972, gw} has emerged as a fundamental benchmark for quantum optimization algorithms due to its dual significance in computational complexity theory and practical applications~\cite{maxcut_app1,maxcut_app2,maxcut_app4_vlsi,maxcut_app5_Wireless,PhysRevApplied.23.014045}. Max-Cut belongs to the class of NP-hard combinatorial optimization problems, making it computationally intractable for classical algorithms to solve optimally on large instances, even when seeking approximate solutions within polynomial time bounds.
Formally, the Max-Cut problem is defined on an undirected graph $G = (V, E)$, where $V$ represents the set of vertices with $|V|$ nodes, and $E$ denotes the set of edges with $|E|$ connections. Each edge $(i,j) \in E$ may be associated with a non-negative weight $w_{ij}$, though the unweighted case where $w_{ij} = 1$ for all edges is commonly studied. A cut $C = (S, \bar{S})$ represents a bipartition of the vertex set $V$ into two disjoint subsets $S$ and $\bar{S} = V \setminus S$. The objective function to maximize is the cut value, defined as $\textsc{CutVal}(C) = \sum_{(i,j) \in E : i \in S, j \in \bar{S}} w_{ij}$, which quantifies the total weight of edges crossing the partition boundary. The Max-Cut problem seeks to identify the partition $(S^*, \bar{S}^*)$ that maximizes this objective function over all possible bipartitions of the vertex set.

Algorithms are typically evaluated using the \emph{approximation ratio} metric, due to the computational complexity inherent in finding exact solutions to Max-Cut instances at scale. This metric quantifies solution quality relative to the optimal value. Specifically, for a given Max-Cut instance with optimal cut value CutVal$_\text{OPT}$ and an algorithm that produces a cut value CutVal$_\text{ALG}$, the approximation ratio is defined as $\text{AR} =$ CutVal$_\text{ALG}$/CutVal$_\text{OPT}$, where $\text{AR} \in [0, 1]$ with higher values indicating better performance. For example, consider a graph where the optimal Max-Cut value is 10, and an optimization algorithm identifies a cut value 9; the corresponding approximation ratio would be $\text{AR} = 9/10 = 0.9$ (or $90$\%). This metric enables meaningful comparison of algorithm performance across different problem instances and scales, particularly when exact optimal solutions are computationally infeasible to determine. A famous classical algorithm for Max-Cut is the Goemans-Williamson (GW) algorithm~\cite{gw}, which achieves a guaranteed approximation ratio of at least 0.878 using semidefinite programming techniques.

Moreover, as an unconstrained discrete optimization problem, Max-Cut admits a natural encoding as a QUBO formulation or equivalently as an Ising model. This property makes the Max-Cut problem well-suited for quantum optimization frameworks, such as the Quantum Approximate Optimization Algorithm.

\subsection{Quantum Approximate Optimization Algorithm}\label{sec:qaoa}

The Quantum Approximate Optimization Algorithm~\cite{QAOA} represents the most prominent quantum heuristic for addressing combinatorial optimization problems on near-term quantum devices, particularly those formulated as QUBO problems. QAOA belongs to the broader class of Variational Quantum Algorithms~\cite{VQAs,VQE}, which leverage hybrid quantum-classical computation paradigms to exploit the complementary strengths of quantum superposition and classical optimization techniques. The algorithm operates through an iterative framework in which a classical optimizer systematically adjusts parameters of a parameterized quantum circuit to maximize the expected value of a problem-specific cost function. This variational approach makes QAOA particularly well-suited for implementation on Noisy Intermediate-Scale Quantum (NISQ) devices, where these quantum processors are characterized by limited number of qubits and susceptibility to noise (that affects gate fidelities). QAOA can tolerate such hardware limitations and still potentially achieve a quantum advantage, thanks to the hybrid quantum-classical computation framework~\cite{qaoa_perf, PhysRevA.103.042412, Barrera2025}.

The mathematical foundation of QAOA rests on the quantum adiabatic theorem and the approximation of adiabatic quantum computation using discrete quantum gates. For a given combinatorial optimization problem encoded as an $n$-qubit cost Hamiltonian $H_C$, QAOA constructs an ansatz quantum state $|\psi({\gamma}, {\beta})\rangle$ through alternating applications of two parameterized unitary operators: the cost operator $U_C(\gamma) = e^{-i\gamma H_C}$ and the mixing operator $U_M(\beta) = e^{-i\beta H_M}$, where $H_M$ is typically chosen as the transverse field Hamiltonian. The ansatz state for $p$ layers is given by $|\psi({\gamma}, {\beta})\rangle = U_M(\beta_p)U_C(\gamma_p)\cdots U_M(\beta_1)U_C(\gamma_1)|+\rangle^{\otimes n}$, where $|+\rangle^{\otimes n}$ is the uniform superposition initial state. The optimization objective is to maximize the expectation value $\langle\psi({\gamma}, {\beta})|H_C|\psi({\gamma}, {\beta})\rangle$, which requires iterative parameter optimization using classical algorithms, such as gradient descent or evolutionary strategies.

The practical implementation of QAOA involves several key factors that influence its performance and scalability~\cite{QAOA_supremacy, BLEKOS20241}. The ansatz circuit depth, set by the number of QAOA layers $p$, introduces a trade-off between solution quality and quantum resource demands: deeper circuits often yield better approximation ratios but require longer coherence times and higher gate fidelity. Parameter initialization strategies also significantly affect solution quality and the number of optimization iterations required~\cite{qaoa_perf, TQA}.

\subsection{Classical Simulation of QAOA}\label{sec:quantum_circuit_simulation}

Quantum circuit simulation~\cite{young2023simulatingquantumcomputationsclassical} on classical computers is an essential tool for quantum algorithm development, given current hardware limitations in qubit count, gate fidelity, and accessibility. Hybrid quantum-classical algorithms, such as QAOA, are often evaluated on classical systems to enable prototyping, theoretical validation, and performance benchmarking on problem instances beyond the reach of existing quantum devices. The computational cost of simulating general quantum circuits grows exponentially with the number of qubits, imposing a fundamental size limitation.
State-vector simulation~\cite{QuEST, cuQuantum,ProjectQ} maintains the complete quantum state and applies gates via matrix-vector multiplications, yielding exact results but requiring $O(2^n)$ memory and computation for $n$ qubits (e.g., 64~GiB for $n = 32$ in double precision), limiting feasibility to relatively small systems. Tensor network approaches~\cite{first_tensornetwork, DMRG}, such as Matrix Product States and related decompositions, exploit limited entanglement to reduce complexity. This allows the simulation of larger quantum systems when entanglement is bounded, but it can come at the cost of accuracy.
Density matrix simulation~\cite{9355323} extends state-vector methods to incorporate quantum noise and decoherence, providing more realistic modeling of NISQ devices with additional computational overhead.

Some existing works have explored the simulation of QAOA circuits, specifically focusing on optimizing the simulation process for the unique structure of QAOA circuits. For instance, the work by Lin et al.~\cite{10.1145/3605098.3635897} presents optimizations on the cost layer and uses multiple GPUs to accelerate the simulation of QAOA circuits. FOR-QAOA~\cite{for_qaoa} is a framework that optimizes the mixer layer of QAOA circuits and allows multi-node parallelization to speed up the simulation process. These works demonstrate that while QAOA circuits can be simulated efficiently on classical hardware, the exponential scaling of quantum state representation remains a fundamental challenge, particularly for larger problem instances.

\subsection{Divide-and-Conquer QAOA for Solving Max-Cut Problems}\label{sec:dc_qaoa}

The divide-and-conquer paradigm in quantum computing can be traced to Dunjko et al.~\cite{dc_quantum_ref1}, who introduced the idea of decomposing large quantum optimization problems into smaller subproblems that can be solved independently. This concept is later extended by Ge et al.~\cite{dc_quantum_ref2}.
In the context of the Quantum Approximate Optimization Algorithm, divide-and-conquer methodologies address the scalability limitations of implementations for large-scale Max-Cut problems. In particular, the divide-and-conquer paradigm partitions the original instance into smaller, manageable subproblems, solves them independently or with limited interdependence, and then combines the solutions to reconstruct the result for the original problem. This strategy mitigates the exponential scaling bottleneck of quantum optimization algorithms, and it enables the solution of Max-Cut instances with hundreds or thousands of vertices on current quantum hardware or classical simulators. Its effectiveness depends on the decomposition scheme, the subproblem-solving method, and the integration procedure used to preserve solution quality.

\begin{figure}[htbp]
    \centering
    \includegraphics[width=\textwidth]{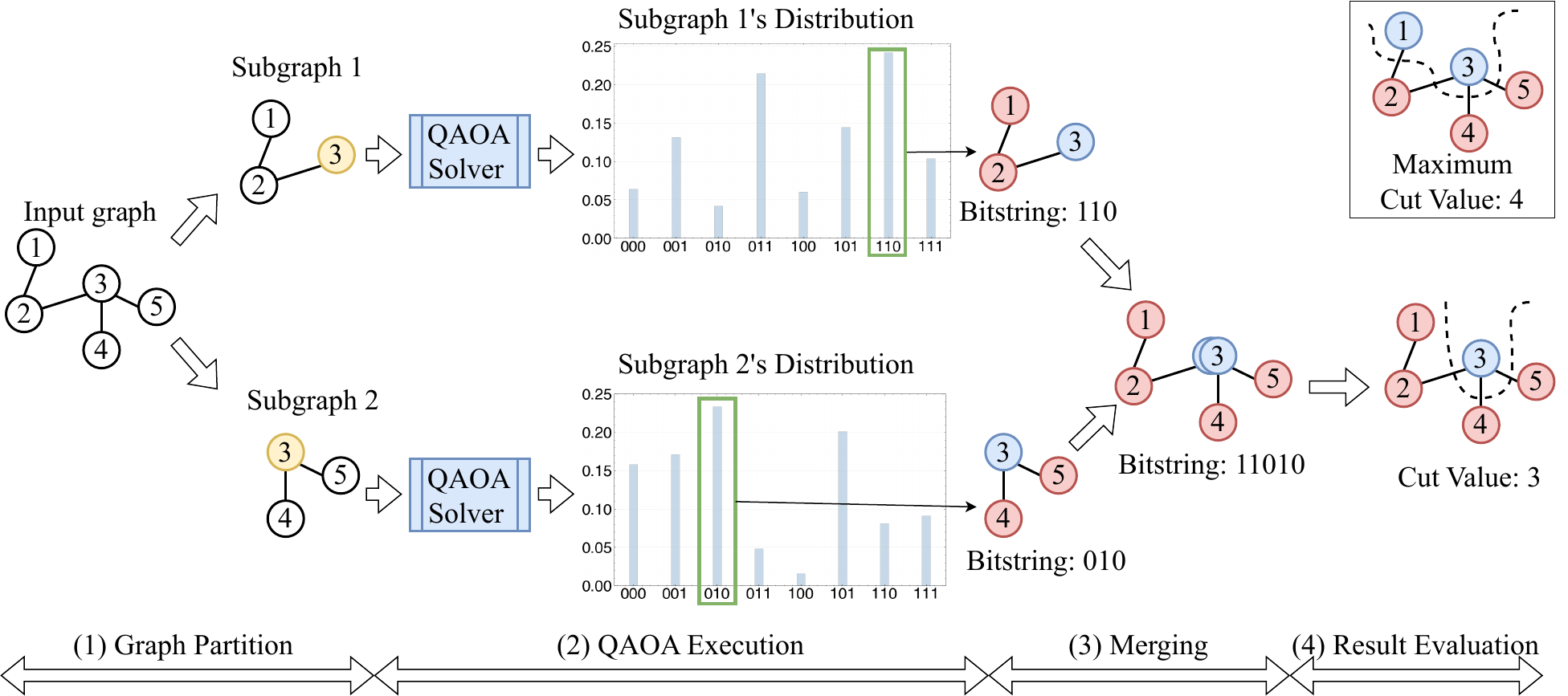}
    \caption{A divide-and-conquer QAOA example for solving the Max-Cut problem.}
    \label{fig:dc_qaoa_example}
\end{figure}

\begin{figure}[htbp]
    \centering
    \includegraphics[width=0.5\textwidth]{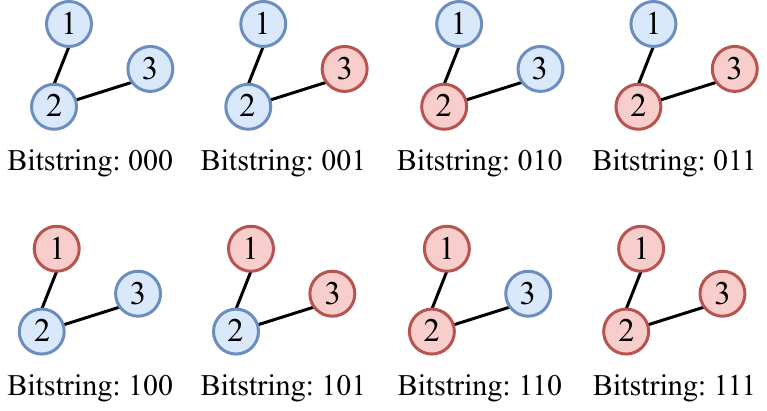}
    \caption{Graphs and their corresponding bitstring representations under different partitions.}
    \label{fig:bitstring_mapping}
\end{figure}

\paragraph{A Divide-and-Conquer QAOA Example}
An example of the divide-and-conquer QAOA approach for solving Max-Cut problems is illustrated in \figurename~\ref{fig:dc_qaoa_example}. The approach consists of four phases.
\begin{enumerate}
    \item \emph{Graph Partition}. The original Max-Cut graph is divided into smaller subgraphs. Some vertices are shared between subgraphs, which are later used in the solution reconstruction phase.
    \item \emph{QAOA Execution}. QAOA is applied independently to each subgraph to obtain approximate solutions. Each subgraph is represented by a \emph{bitstring} that encodes its partition scheme, with the mapping illustrated in \figurename~\ref{fig:bitstring_mapping}. For example, subgraph~1 yields the bitstring \texttt{110}, indicating vertices $v_1$ and $v_2$ belong to set $S$ and $v_3$ belongs to $\bar{S}$.
    \item \emph{Merging}. Subgraph solutions are combined by concatenating their bitstrings. This step requires consistent assignments for shared vertices. For instance, in \figurename~\ref{fig:dc_qaoa_example}, the two subgraphs can only be merged if both assign $v_3$ to the same set. As the vertex $v_3$ in the bitstrings for subgraphs 1 and 2 is \texttt{0}, and the two subgraphs can be merged to form the bitstring \texttt{11010}.
    \item \emph{Result Evaluation}. The merged bitstring is evaluated against the original Max-Cut objective to maximize the number of edges between $S$ and $\bar{S}$. As illustrated in the top-right of \figurename~\ref{fig:dc_qaoa_example}, the resulting solution may not be optimal (with the cut value of 3). Therefore, the approximation ratio is used to assess solution quality, and it is defined as the ratio of the cut value from the merged bitstring to the optimal cut value of the original graph (e.g., 3/4 in the example).
\end{enumerate}

\paragraph{Existing Graph Partitioning Strategies}

Graph partitioning is a key preprocessing step in divide-and-conquer approaches to combinatorial optimization, including Max-Cut. Its objective is to decompose a graph into smaller subgraphs while minimizing inter-partition edges. Minimizing these edges reduces subproblem complexity, enables parallel execution, and preserves structural properties that support high-quality solutions.
Traditional partitioning algorithms often have high computational complexity, scaling quadratically or worse with the number of vertices. For large instances, this can create significant preprocessing bottlenecks, particularly in frameworks that rely on sophisticated partitioning to achieve strong approximation ratios.
In quantum optimization, especially within divide-and-conquer based frameworks, partitioning enables large-scale Max-Cut instances to fit within hardware limits, such as qubit capacity. While dividing the graph into smaller subgraphs reduces computational demands, removing inter-partition edges can degrade AR. This creates a trade-off between computational efficiency and solution quality.
Existing strategies vary in their trade-offs. DC-QAOA~\cite{dcqaoa} uses a Large Graph Partitioning (LGP) algorithm to achieve strong AR by balancing subgraphs, but its high runtime and connectivity constraints limit scalability. QAOA$^2$~\cite{QiQ} employs randomized partitioning for faster execution. It offsets potential AR loss through advanced merging strategies, though this introduces additional merging overhead. Coupling QAOA~\cite{CQ} adopts a binary decomposition that preserves connectivity via coupling terms in the Hamiltonian, but it is restricted to binary partitions and does not scale effectively to larger graphs.

\paragraph{Limitations of Existing Divide-and-Conquer QAOA Approaches}

Existing divide-and-conquer QAOA variants, such as DC-QAOA~\cite{dcqaoa}, QAOA$^2$~\cite{QiQ}, and Coupling QAOA~\cite{CQ}, primarily focus on solution quality but face significant limitations that reduce their practical utility and scalability. These limitations can be grouped into three main categories.
First, \emph{execution time bottlenecks} arise from computationally expensive preprocessing, particularly graph partitioning algorithms with quadratic or higher time complexity in the number of vertices. For example, DC-QAOA relies on sophisticated partitioning with $O(|V|^2)$ or higher complexity, which can dominate total runtime for large instances and offset potential quantum speedups. Since DC-QAOA's implementation is not publicly available, our evaluation is based on a reimplementation from their description. QAOA$^2$ and Coupling QAOA incur additional overhead from exhaustive searches over subproblem combinations, leading to exponential scaling in both runtime and memory usage.
Second, \emph{parameter optimization} is often ad hoc or absent, resulting in convergence issues and inconsistent solution quality. Without systematic parameter initialization and tuning, performance varies significantly across problem instances, limiting reliability for practical deployment.
Third, \emph{scalability constraints} limit the extent of problem decomposition and the maximum problem size addressable. Coupling QAOA's binary-only decomposition restricts problem size reduction. DC-QAOA suffers performance degradation on dense graphs, where partitioning is difficult. QAOA$^2$ lacks efficient parallelization for subproblem execution, reducing scalability gains.
Chuang et al.~\cite{10674517} also identified these challenges and emphasized the need for a more comprehensive, systematically designed approach. However, their evaluation was limited to small-scale Max-Cut instances (up to 26 vertices), insufficient to assess the scalability of divide-and-conquer QAOA methods.

\subsection{Motivation}\label{sec:motivation}

As elaborated in \sectionname~\ref{sec:dc_qaoa}, current divide-and-conquer QAOA methods favor solution quality over execution efficiency. This design choice limits their effectiveness for large or complex problem instances. For problems exceeding 100 vertices, evaluation becomes challenging due to substantial variation in execution time. From our experiments, QAOA$^2$ requires approximately 4.7 hours to compute a solution for a medium-scale, high-density graph with 400 vertices, and is estimated to take about 13.6 days for a large-scale, high-density graph with 16,000 vertices. Further results are presented in \sectionname~\ref{sec:evaluation}. The key observation is that prior work develops novel graph partitioning algorithms to lower the computation load for the subgraphs (e.g., LGP in DC-QAOA and binary decomposition in Coupling QAOA) and employs intricate merging strategies to mitigate approximation ratio degradation, but these come at the cost of increased execution time. Unfortunately, these approaches do not fully leverage the presence of high performance computing hardware, such as multi-core processors and multiple GPUs, which can be leveraged to parallelize the execution of subproblems and solution reconstruction. As a result, their scalability and efficiency for large Max-Cut problems remain limited.

By recognizing their limitations and considering the presence of high-performance computing hardware, we propose \emph{ParaQAOA}, a framework that incorporates an efficient graph partitioning algorithm, a parallelized execution flow, a systematic parameterized design, and a performance evaluation metric for Max-Cut problems. The features of these innovations are highlighted as follows.
First, a linear-time graph partitioning algorithm is developed. It reduces decomposition complexity from quadratic or higher to linear in the number of vertices by exploiting structural properties. This improvement eliminates preprocessing bottlenecks and enables efficient partitioning of large graphs. Second, a fully parallelized execution flow is introduced. This parallel execution pipeline encompasses both subproblem solving and solution reconstruction and enables effective use of modern parallel architectures to reduce overall runtime.
Third, a systematic parameterized design is proposed. This design incorporates hardware platform specifications for parameter configuration and leaves tunable parameters for users to manage the trade-off between execution efficiency and solution quality. These parameters are introduced during the framework's introduction in the following section, and a concrete example of how to set up these parameters is introduced in \sectionname~\ref{sec:effectiveness_of_parameter_configurations}.
Fourth, a unified evaluation metric is introduced. This metric jointly considers solution quality and execution efficiency, enabling consistent comparisons and informed trade-off decisions across different algorithms and configurations. This is particularly important for large-scale problems, where significant variations in execution time can complicate the evaluation of algorithm effectiveness.
Together, these innovations enable ParaQAOA to achieve superior scalability and computational efficiency for addressing large-scale Max-Cut problems.
\section{ParaQAOA}\label{sec:method}

In this section, we present our methodology for solving large-scale Max-Cut problems. The overview of the proposed ParaQAOA framwork is presented in \sectionname~\ref{sec:frameworkoverview}. After that, three key components of the ParaQAOA framework are described in detail: the graph partitioning strategy in \sectionname~\ref{sec:graphpartition}, the parallelized QAOA execution in \sectionname~\ref{sec:qaoaexecution}, and the level-aware parallel merge process in \sectionname~\ref{sec:parallelmerge}. Finally, the proposed Performance Efficiency Index is introduced in \sectionname~\ref{subsec:pei_framework} to systematically evaluate the performance trade-offs for methodologies used to handle large-scale Max-Cut problems.

\subsection{Framework Overview}\label{sec:frameworkoverview}

\begin{figure}[htbp]
    \centering
    \includegraphics[width=1.0\textwidth]{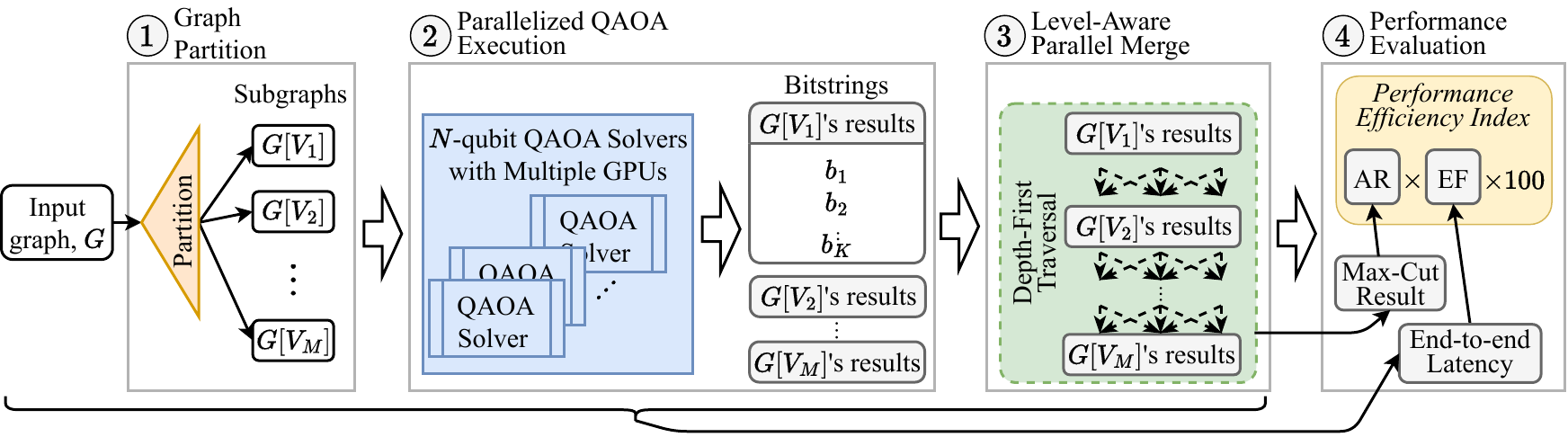}
    \caption{ParaQAOA framework overview.}
    \Description{Block diagram showing the main components of the ParaQAOA architecture, including graph partition, QAOA execution, and parallel merge.}
    \label{fig:adc_qaoa_arch}
\end{figure}

The proposed ParaQAOA framework employs a hierarchical and highly parallel architecture to efficiently solve large-scale Max-Cut problems by leveraging quantum computing capabilities. The architecture addresses the computational complexity of large graphs through decomposition into manageable subgraphs, parallel processing, and structured result aggregation.
The framework incorporates several key optimizations to enhance computational efficiency and improve solution quality in solving the Max-Cut problem. As illustrated in \figurename~\ref{fig:adc_qaoa_arch}, these optimizations are organized into three sequential stages to progress from the input graph to the final output solution. Additionally, a performance evaluation stage is included to generate the proposed Performance Efficiency Index, providing a quantitative measure for assessing the efficiency of the obtained Max-Cut solutions. The key components of the framework are described as follows.

\begin{enumerate}[label=\protect\circleda{\arabic*}]
    \item The \textbf{Graph Partition} stage decomposes the input graph $G$ into multiple smaller subgraphs $\{G[V_i]\}_{i=1}^M$, where each pair of adjacent subgraphs shares at most one node. This partitioning ensures that problem sizes remain manageable for the QAOA solvers, which typically have constraints on the number of qubits they can handle. While various partitioning strategies are available, in this framework, we employ a \emph{Connectivity-Preserving Partitioning} Algorithm that efficiently divides the graph into a predefined number of subgraphs with approximately equal sizes while preserving essential connectivity information. This approach enables effective load balancing across the available quantum resources, thereby facilitating scalable parallel processing of large graphs.

    \item The \textbf{Parallelized QAOA Execution} stage assigns each subgraph to a QAOA solver operating concurrently across multiple computational units, such as GPUs, to efficiently explore the solution space of each subgraph. The subgraphs $\{G[V_i]\}_{i=1}^M$  will be batched into multiple QAOA Solvers. For each subgraph, the QAOA solver outputs a set of top-$K$ bitstrings $\{b_i\}_{i=1}^K$, where $b_i {\in} \{0,1\}^{|V_i|}$ sorted by their corresponding probability $\pi_i$; that is, each $b_i$ is a binary string of length $|V_i|$ representing a bipartition of the vertex set $V_i$. Moreover, we proposed a \emph{Selective Distribution Exploration Strategy} that provides flexibility in adjusting the number of bitstrings $K$ to balance computational resources and solution quality. This stage leverages parallel execution for efficiently exploring the solution space of each subgraph, and fexibility in adjusting the number of bitstrings to balance computational resources and solution quality.

    \item The \textbf{Level-Aware Parallel Merge} stage reconstructs the bitstring results from all subgraphs using depth-first traversal of the Cartesian product space from the bitstring set $\{b_i\}_{i=1}^K$ to enumerate possible global solutions. For each reconstructed global bitstring, the potential Max-Cut value is calculated again by considering all edges of the original input graph $G$. The algorithm then selects the bitstring yielding the maximum cut value as the final output. Moreover, this methodology provides dynamic flexibility by adjusting the candidate pool size to make balance between computational efficiency and solution quality.

    \item The \textbf{Performance Evaluation} stage systematically evaluates the effectiveness of the large-scale Max-Cut solving for balancing execution time and approximation ratio. We introduce a new performance index, \emph{Performance Efficiency Index}, a composite metric designed to quantify the fundamental trade-off between solution quality and computational efficiency for benchmarking different Max-Cut problem-solving approaches.
\end{enumerate}

The following subsections detail our implementation of each component within the ParaQAOA framework.

\subsection{Graph Partition}\label{sec:graphpartition}
\begin{figure}[htbp]
    \centering
    \includegraphics[width=0.95\columnwidth]{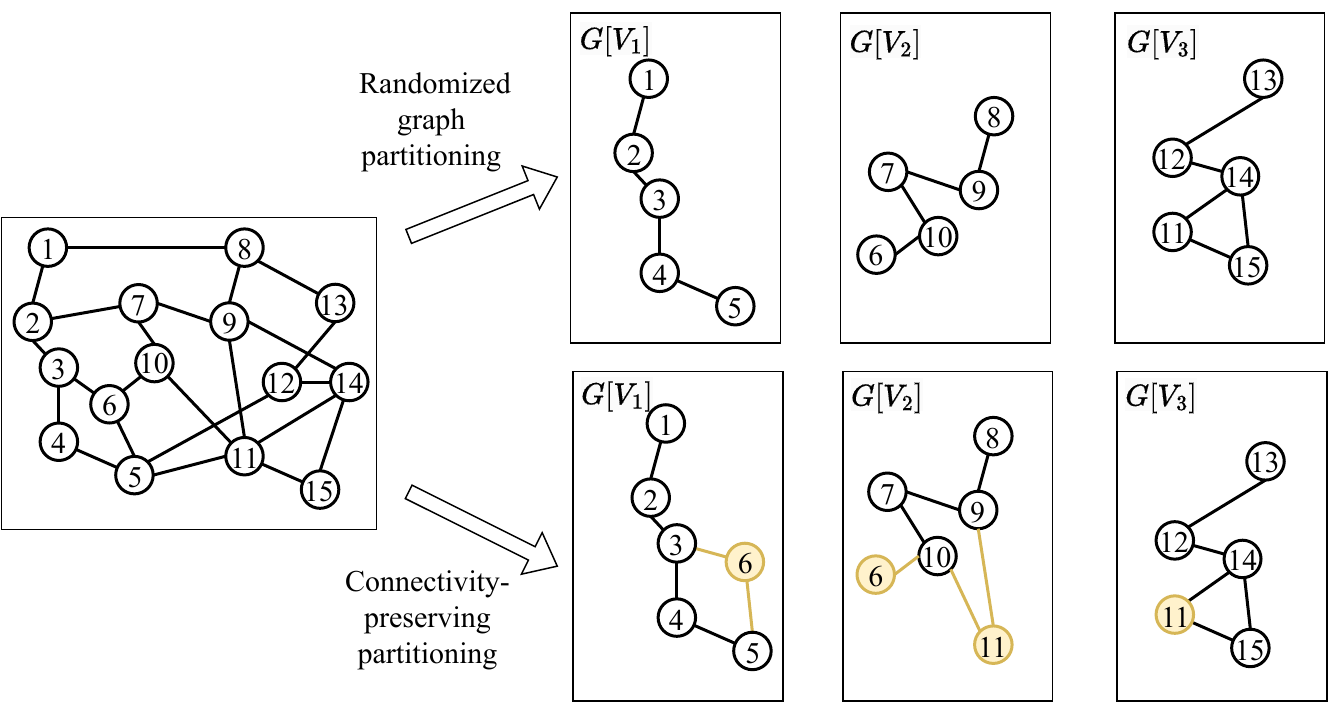}
    \caption{Comparing random partitioning and proposed connectivity-preserving partitioning algorithms.}
    \label{fig:partitioning}
\end{figure}

The graph partitioning phase is a critical component of DC-QAOA-like frameworks and enables efficient parallel processing of large-scale Max-Cut problems by decomposing the input graph into smaller subgraphs. We improve upon the DC-QAOA~\cite{dcqaoa} and QAOA$^2$~\cite{QiQ} algorithms by proposing the \emph{Connectivity-Preserving Partitioning} algorithm, which divides the input graph $G$ into $M$ subgraphs $\{G[V_1], G[V_2], \ldots, G[V_M]\}$. The number of subgraphs $M$ is determined based on the capacity of the available QAOA solvers and the underlying hardware resources. Most importantly, each subgraph preserves essential connectivity information, as illustrated in \figurename~\ref{fig:partitioning}, while reducing the problem size to fit within the constraints of an $N$-qubit QAOA solver.

\subsubsection{Partitioning Constraints}
Given an input graph $G=(V,E)$ with $|V|$ vertices, we seek to partition $G$ into $M$ subgraphs while maintaining computational feasibility for subsequent QAOA processing. The partitioning function $P: G \rightarrow \{{G[V_i]}\}_{i=1}^M$ satisfies the following constraints.

\begin{enumerate}
    \item Adjacent subgraph connectivity: $|V_i \cap V_{i+1}| = 1$ for all $i \in \{1,2,\ldots,M{-}1\}$, meaning that subgraph $i$ and subgraph $i+1$ share exactly one node. This property creates a sequential chain of connected subgraphs to preserve connectivity information.
    \item QAOA compatibility: $|V_i| \leq N$ for each subgraph $G[V_i]$, where $N$ denotes the number of qubits available in the QAOA solver. This property ensures that each subgraph can be processed by the QAOA solver without exceeding the solver's qubit capacity.
    \item Size balancing: $|V_i| \leq \left\lceil {|V|/M} \right\rceil$ for each subgraph $G[V_i]$, ensuring a balanced distribution of vertices across partitions. A simpler approach is to set $|V_i| = \left\lfloor {|V|/M} \right\rfloor$ for each subgraph, which satisfies the constraint. Partitioning the graph into subgraphs of similar size ensures balanced computational loads.
\end{enumerate}

\subsubsection{Partitioning Algorithm}
We leverage an efficient \emph{Connectivity-Preserving Partitioning} approach to achieving a computational complexity of $O(|V| {+} |E|)$ while maintaining acceptable solution quality. The algorithm takes as input the graph $G = (V, E)$ and the desired number of subgraphs $M$, and outputs $M$ subgraphs $(G[V_1], G[V_2], \ldots, G[V_M])$. The core method divides the vertices of $G$ into $M$ approximately equal-sized groups and then generates an induced subgraph for each group, preserving all original connections between vertices within the same group.

Algorithm~\ref{alg:random_graph_partition} details the implementation of our efficient graph partitioning approach. First, the base partition size is calculated as $s = \left\lfloor {|V|/M} \right\rfloor {-} 1$ to guarantee a balanced distribution of vertices among partitions. The subtraction of $1$ accounts for the shared node to be added later. The algorithm then iterates from $1$ to $M$, performing graph partitioning while tracking vertex assignments to each of the $M$ partitions and employing a node-sharing technique to preserve connectivity information across subgraphs. The last partition accommodates any remaining vertices to handle cases where $|V|$ is not perfectly divisible by $M$. For each partition, the subgraph $G[V_i]$ is generated using $\textsc{GetSubgraph}$, which collects all edges $(u,v) \in E$ where both endpoints belong to the same vertex set $V_i$. Finally, the list of partitioned subgraphs, $\{G[V_i]\}_{i=1}^M$, is returned.

\begin{algorithm}[htbp]
    \caption{Connectivity-Preserving Partitioning: a efficient graph partitioning that preserves connectivity information of adjacent subgraphs through shared nodes.}
    \label{alg:random_graph_partition}
    \KwIn{$G$, input graph,\newline
        $M$, number of subgraphs}
    \KwOut{$\{G[V_1], G[V_2], \ldots, G[V_M]\}$, partitioned subgraphs }

    $V \leftarrow V(G)$\Comment*[r]{set of vertices in $G$, indexed from $0$ to $|V|-1$}
    $s \leftarrow \lfloor |V|/M \rfloor - 1$\Comment*[r]{load-balance partition size; `$-1$' is for node sharing later}
    $\text{subgraphs} \leftarrow \emptyset$\;

    \For{$i \leftarrow 1$ \KwTo $M$}{
        $\text{start\_index} \leftarrow (i{-}1) \times s$\;
        \eIf{$i = M$}{
            $\text{end\_index} \leftarrow |V|$\;
        }{
            $\text{end\_index} \leftarrow \text{start\_index} + s + 1$\Comment*[r]{`$+1$' is for node sharing}
        }
        $V_i \leftarrow \{v \in V \mid \text{start\_index} \leq \text{index}(v) < \text{end\_index}\}$\;
        $G[V_i] \leftarrow$ \GetSubgraph{$G$, $V_i$}\;
        $\text{subgraphs} \leftarrow \text{subgraphs} \cup \{G[V_i]\}$\;
    }
    \Return{subgraphs}\;
    \BlankLine\BlankLine
    \Fn{\GetSubgraph{$G$, $V_i$}}{
        Initialize $G[V_i] = (V_i, E_i)$ where $E_i = \emptyset$\;
        \ForEach{$(u, v) \in E$}{
            \If{$u \in V_i$ \textbf{and} $v \in V_i$}{
                $E_i \leftarrow E_i \cup \{(u, v)\}$\;
            }
        }
        \Return{$G[V_i]$}\;
    }
\end{algorithm}

While this approach may appear simplistic compared to sophisticated partitioning schemes, experimental results demonstrate its effectiveness in maintaining solution quality while enabling rapid processing of large-scale instances. The method is particularly well-suited for dense graphs where edge distribution is approximately uniform.
The primary advantages of this methodology are: 1) linear-time preprocessing suitable for large-scale instances, 2) deterministic execution time independent of graph structure, 3) minimal memory usage, and 4) excellent scalability properties.
These characteristics make our partitioning approach particularly suitable for solving large-scale Max-Cut problems within the ParaQAOA framework, where computational efficiency is paramount.

\subsection{Parallelized QAOA Execution}\label{sec:qaoaexecution}

\begin{figure}[htbp]
    \centering
    \includegraphics[width=\columnwidth]{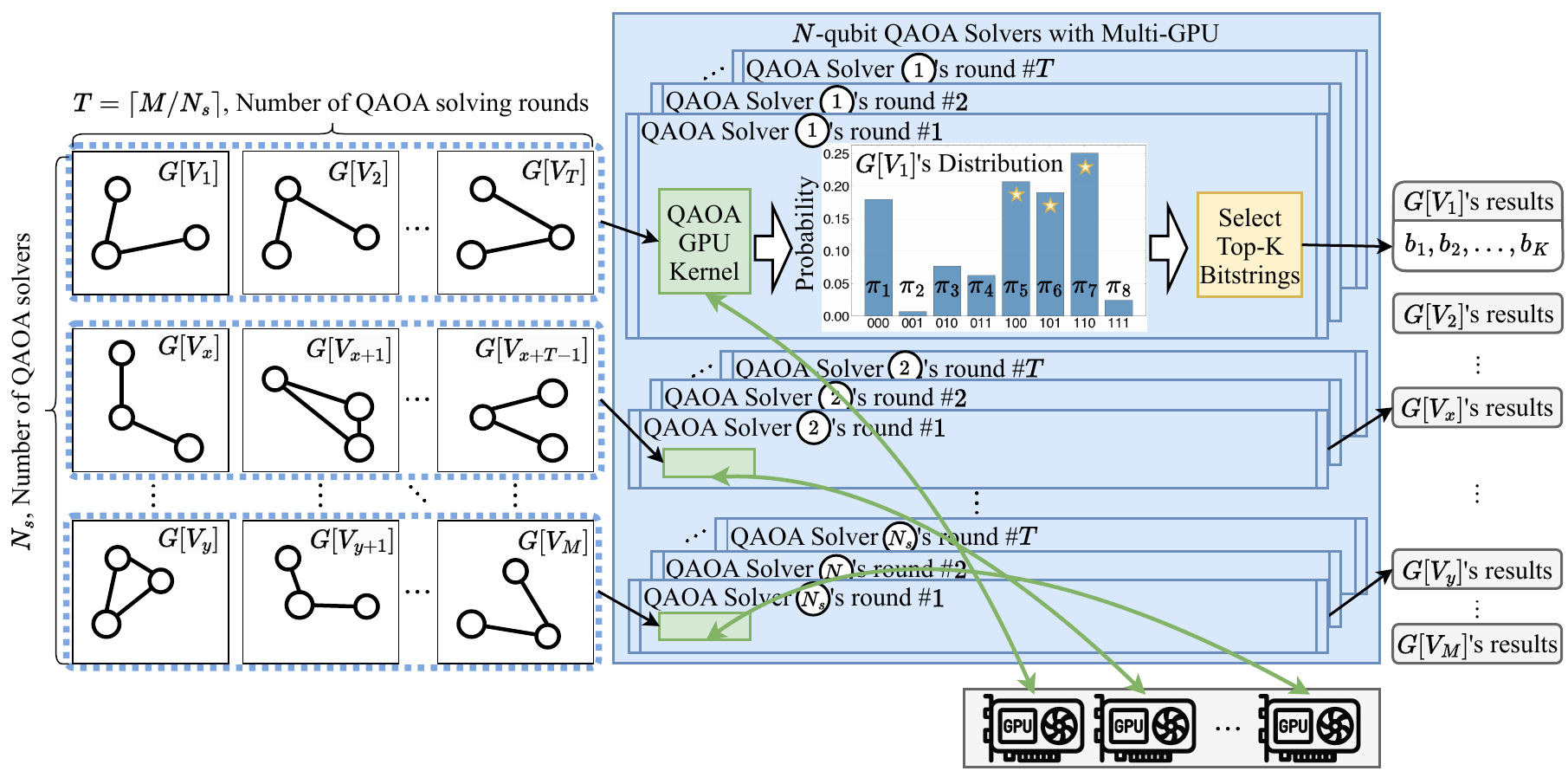}
    \caption{The proposed parallelized QAOA execution workflow integrated with a multi-GPU system. }
    \Description{Diagram illustrating the workflow of the QAOA solving process, including subgraph processing, GPU execution, and result aggregation.}
    \label{fig:qaoa_execution}
\end{figure}

The parallelized QAOA execution phase is the main computational stage of our framework, in which simultaneous quantum computation is performed across all partitioned subgraphs. As shown in \figurename~\ref{fig:qaoa_execution}, this phase takes the $M$ subgraphs generated by the Graph Partition phase and executes QAOA on each subgraph independently.
For each subgraph $G[V_i]$, the execution produces a set of top-$K$ bitstrings $\{b_1, b_2, \ldots, b_K\}$, where each bitstring encodes a vertex assignment (0 or 1) and is selected based on its high probability in the QAOA output distribution. These filtered bitstrings serve as candidate solutions for the subsequent merge phase.

Our framework employs an $N$-qubit QAOA Solver Pool, where $N$ is determined based on the available quantum processing units (QPUs) or classical computing resources. We utilize a \emph{Multi-GPU Computing Architecture} to execute QAOA in parallel, with each subgraph assigned to one of the QAOA solvers. Each solver performs quantum circuit execution to calculate the solution distribution for its assigned subgraph.
Additionally, we provide a \emph{Selective Distribution Exploration Strategy} by making the parameter $K$ be configurable by the user and can be adjusted to potentially increase the approximation ratio (with a larger $K$ value). That is, the flexibility in balancing computational resources against solution quality is controlled by the parameter $K$.

\subsubsection{Multi-GPU Computing Architecture}
Simulating quantum circuits for multiple subgraphs requires an effective distributed computing approach. Our multi-GPU architecture addresses this challenge through dynamic resource allocation and parallel execution. As shown in the middle of \figurename~\ref{fig:qaoa_execution}, the system uses a load balancing strategy to distribute subgraphs across GPUs based on subgraph size and current GPU utilization, ensuring optimal resource use and minimizing idle time.
This parallel execution framework processes multiple quantum circuits at the same time, with each subgraph computed independently. This design achieves near-linear speedup as the number of GPUs increases, which is important for large problem instances where sequential processing would be too slow. Synchronized result aggregation ensures that all QAOA executions finish before proceeding to the next phase.

In the current prototype implementation, we utilize the QAOA GPU kernel from Lu et al.~\cite{CQ}, which employs the \emph{numba}~\cite{numba} package to accelerate QAOA execution on NVIDIA GPUs. This kernel computes the full quantum state distribution for each subgraph, where each bitstring corresponds to a potential solution. It efficiently applies quantum gates based on the subgraph's Hamiltonian and outputs the resulting bitstring probabilities by leveraging GPU parallelism.

As illustrated in \figurename~\ref{fig:qaoa_execution}, given $N_s$ available QAOA solvers and $M$ subgraphs, execution proceeds in $T = \lceil M / N_s \rceil$ rounds. In each round, a QAOA GPU kernel processes a subgraph and outputs a probability distribution $\{\pi_1, \pi_2, \ldots, \pi_{2^N}\}$ over $2^N$ bitstrings, where $N$ is the number of qubits in the subgraph.
For example, the subgraph $G[V_1]$ is processed in the first round, producing a distribution $\{\pi_1^{(1)}, \pi_2^{(1)}, \ldots, \pi_8^{(1)}\}$. This process is repeated for all subgraphs, with each QAOA solver outputting candidate solutions to the Max-Cut problem in the form of bitstrings and their associated probabilities.

\subsubsection{Selective Distribution Exploration Strategy}

A QAOA solver generates $2^N$ bitstrings per subgraph, where $N$ is the number of qubits corresponding to the subgraph size. To reduce computational complexity and focus on high-quality candidates, we employ a \emph{Selective Distribution Exploration Strategy} that filters bitstrings based on their measured probabilities. As a result, each QAOA execution yields a set of $K$ bitstrings $\{b_1^{(i)}, b_2^{(i)}, \ldots, b_K^{(i)}\}$ for the $i$-th subgraph, representing the most promising solutions identified by quantum optimization.

Each bitstring $b_j^{(i)} \in \{0,1\}^{|V_i|}$ encodes a binary partitioning of subgraph $G[V_i]$, where $|V_i|$ is the number of vertices and $1 \leq j \leq K$. The value $b_j^{(i)}[v] = 0$ assigns vertex $v$ to one partition, and $b_j^{(i)}[v] = 1$ to the other. While high-probability bitstrings typically correspond to better cuts, the strategy also preserves diversity by allowing lower-probability candidates, so as to acknowledge that globally optimal solutions may reside in less probable regions of the state space.

The parameter $K$ controls the number of candidate solutions retained for each subgraph after QAOA execution. A larger $K$ increases the diversity of solutions considered in the subsequent merge phase, potentially improving the approximation ratio but incurring higher computational overhead. Conversely, a smaller $K$ limits the selection to top-performing bitstrings, reducing complexity while focusing on the most probable solutions.
As the simplified example illustrated in \figurename~\ref{fig:qaoa_execution}, when $K = 3$, the ParaQAOA framework selects the top-3 bitstrings with the highest probabilities, e.g., $\{b_1^{(1)}, b_2^{(1)}, b_3^{(1)}\} = \{100, 101, 110\}$ for subgraph $G[V_1]$. These bitstrings are then stored as candidate results for use in the merge phase.

This tunable parameter enables a trade-off between computational efficiency and solution quality, allowing users to tailor the framework to different application constraints. Although the theoretical solution space for each subgraph consists of $2^{|V_i|}$ configurations, QAOA naturally concentrates probability mass on a small subset of high-quality solutions, efficiently identifying promising candidates through quantum superposition.

\subsection{Level-Aware Parallel Merge}\label{sec:parallelmerge}

The level-aware parallel merge phase is responsible for reconstructing global solutions from the parallel QAOA execution results. While this phase does not impact the approximation ratio, it constitutes the most computationally intensive post-processing step to calculate and find the Max-Cut from plenty of potential solutions. Substantial hardware resources are required for optimal performance. The merge process reconstructs problem solutions by merging (concatenating) subgraph results to identify configurations that maximize the global cut value for the original graph. The concatenation process employs \emph{Parallel Depth-First Traversal Merging} of the Cartesian product space formed by subgraph results, where each level corresponds to one subgraph's solution set and branches represent potential solution paths through the combinatorial space.
Moreover, to fully leverage parallel processing capabilities, the merge phase is designed to operate in a \emph{Level-Aware Scheme}, enabling the use of additional hardware resources to efficiently perform the merging process and identify the maximum cut result.

\subsubsection{Parallel Depth-First Traversal Merging}

\begin{figure}
    \centering
    \includegraphics[width=\columnwidth]{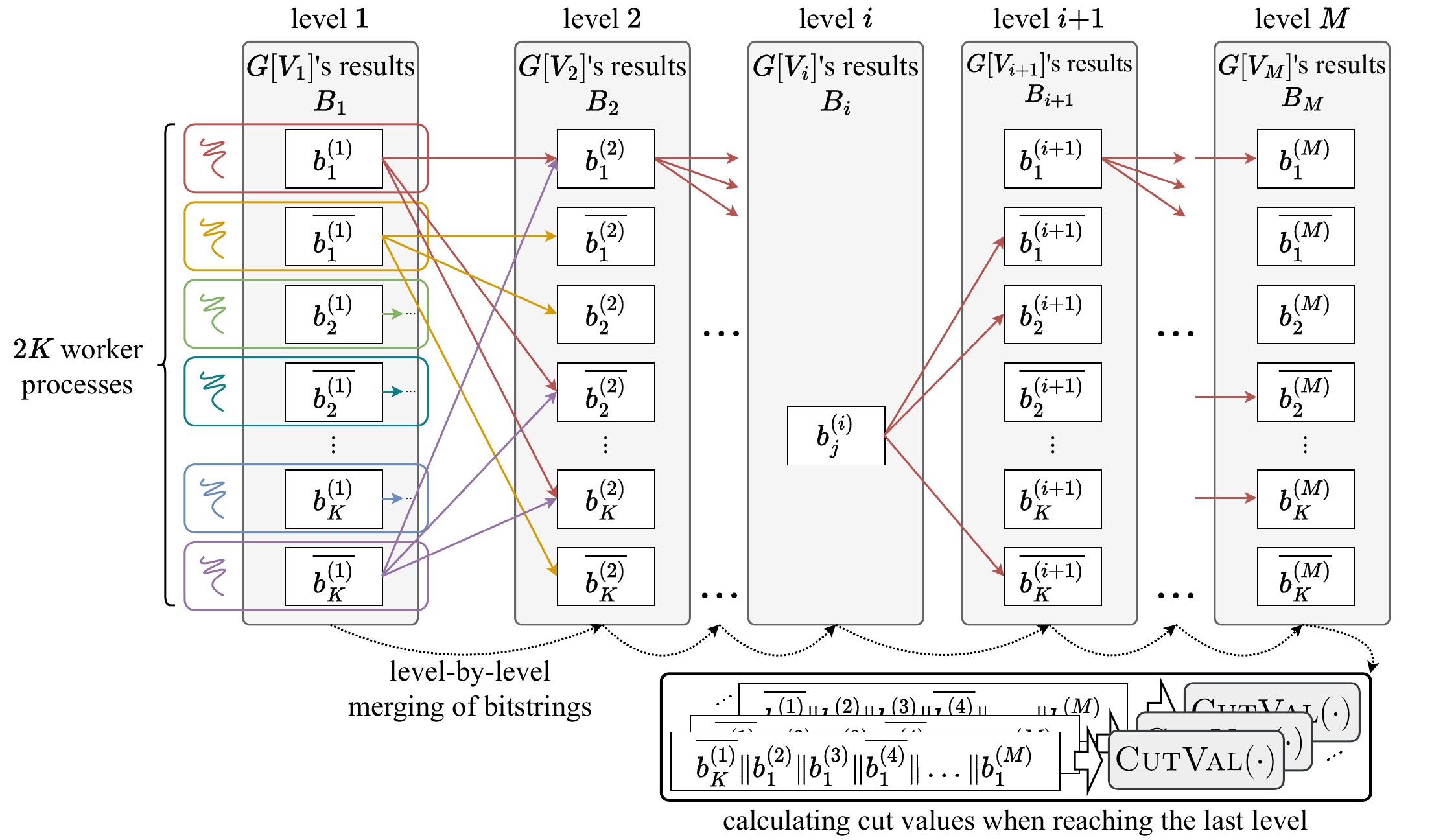}
    \caption{Parallel depth-first traversal merging workflow. Multiple worker processes explore the solution space by starting from different bitstring results of $G[V_1]$ and traversing the Cartesian product space formed by the subgraph solutions.}
    \label{fig:level_aware_merge1}
\end{figure}

The merge phase workflow is illustrated in \figurename~\ref{fig:level_aware_merge1}, where the results from multiple subgraphs are presented in a level-wise manner and combined to form global candidate solutions. 
Each level corresponds to a subgraph, and each branch represents a potential solution path through the combinatorial space.
This process uses a parallel depth-first traversal of the solution space formed by the subgraph solution sets, with each worker process assigned to explore a specific branch starting from the bitstrings of $G[V_1]$. When the traversal reaches the final level, corresponding to the total number of subgraphs (as shown in the bottom of \figurename~\ref{fig:level_aware_merge1}), the $\textsc{CutVal}(\cdot)$ function is invoked to calculate the cut value for the concatenated bitstring. The algorithm then updates the current maximum cut value if a better solution is found.

The total candidate solution space for the merging stage is defined as the Cartesian product of the candidate bitstring sets from each subgraph. Let $B_i = \{ b_1^{(i)}, \overline{b_1^{(i)}}, b_2^{(i)}, \overline{b_2^{(i)}}, \ldots, b_j^{(i)}, \overline{b_j^{(i)}}, \ldots, b_K^{(i)}, \overline{b_K^{(i)}} \}$ denote the set of candidate bitstrings for subgraph $i$, where $\overline{b}$ represents the bitwise inverse of $b$, capturing both possible group assignments for each solution. The Cartesian product space is then defined as:
\[
    B_1 \times B_2 \times \cdots \times B_M
    = \left\{ (b^{(1)}, b^{(2)}, \ldots, b^{(M)}) \mid b^{(i)} \in B_i \text{ for each } i=1,2,\ldots,M \right\}.
\]
This Cartesian product space represents the total set of candidate global solutions formed by selecting one bitstring from each subgraph for evaluation in the merging stage. Although each $B_i$ contains $2K$ bitstrings to account for both original and inverted representations, only \emph{half} can be selected at each level due to the connectivity constraints preserved during partitioning. Specifically, for any bitstring $b_j^{(i)}$ from subgraph $i$, the possible concatenation at the next level is either $b_{j^\prime}^{(i+1)}$ or its inverse $\overline{b_{j^\prime}^{(i+1)}}$. Therefore, the total number of candidate solutions is $2K^M$, where $K$ is the number of top bitstrings retained by the QAOA solvers and $M$ is the number of subgraphs.

Finally, the algorithm updates the maximum cut value by identifying the optimal global configuration, $B^*$. Formally, given $M$ subgraphs $\{ G[V_1], G[V_2], \ldots, G[V_M] \}$ with corresponding QAOA solution sets $\{ B_1, B_2, \ldots, B_M \}$, the merge phase seeks:
\[
    B^* = b^{(1)} \Vert b^{(2)} \Vert \cdots \Vert b^{(M)}, \text{where } b^{(i)} \in B_i \text{ for each } i=1,2,\ldots,M,
\]
to maximize the global cut value:
\[
    \text{Cut}(B^*) = \sum_{i=1}^{M} \text{Cut}(b^{(i)}) + \sum_{(u,v) \in E_{\text{inter}}} \mathbb{I}\big[ B^*[u] \neq B^*[v] \big],
\]
where $E_{\text{inter}}$ denotes the set of inter-partition edges discarded during partitioning, and $\mathbb{I}[\cdot]$ is the indicator function. The \emph{indicator function $\mathbb{I}[\cdot]$} is a mathematical function that returns $1$ if the condidition inside the brackets is true, and $0$ otherwise. In this context, it evalutes each inter-partition edge $(u,v)$ by checking whether the vertices $u$ and $v$ are assigned to different partitions ($0$ or $1$). in the reconstructed global solution $B^*$. If $B^{*}[u] \neq B^{*}[v]$, the edge contributes to the cut and the function returns $1$; otherwise, it returns $0$. This mechanism effectively ensures that the totla cut value accounts for both the intra-subgraph cuts calculated by individual QAOA solvers and the inter-partition edges that were discarded during the partitioning phasembut must be considerd in the final global solution.

\subsubsection{Level-Aware Scheme}

\begin{figure}
    \centering
    \includegraphics[width=\columnwidth]{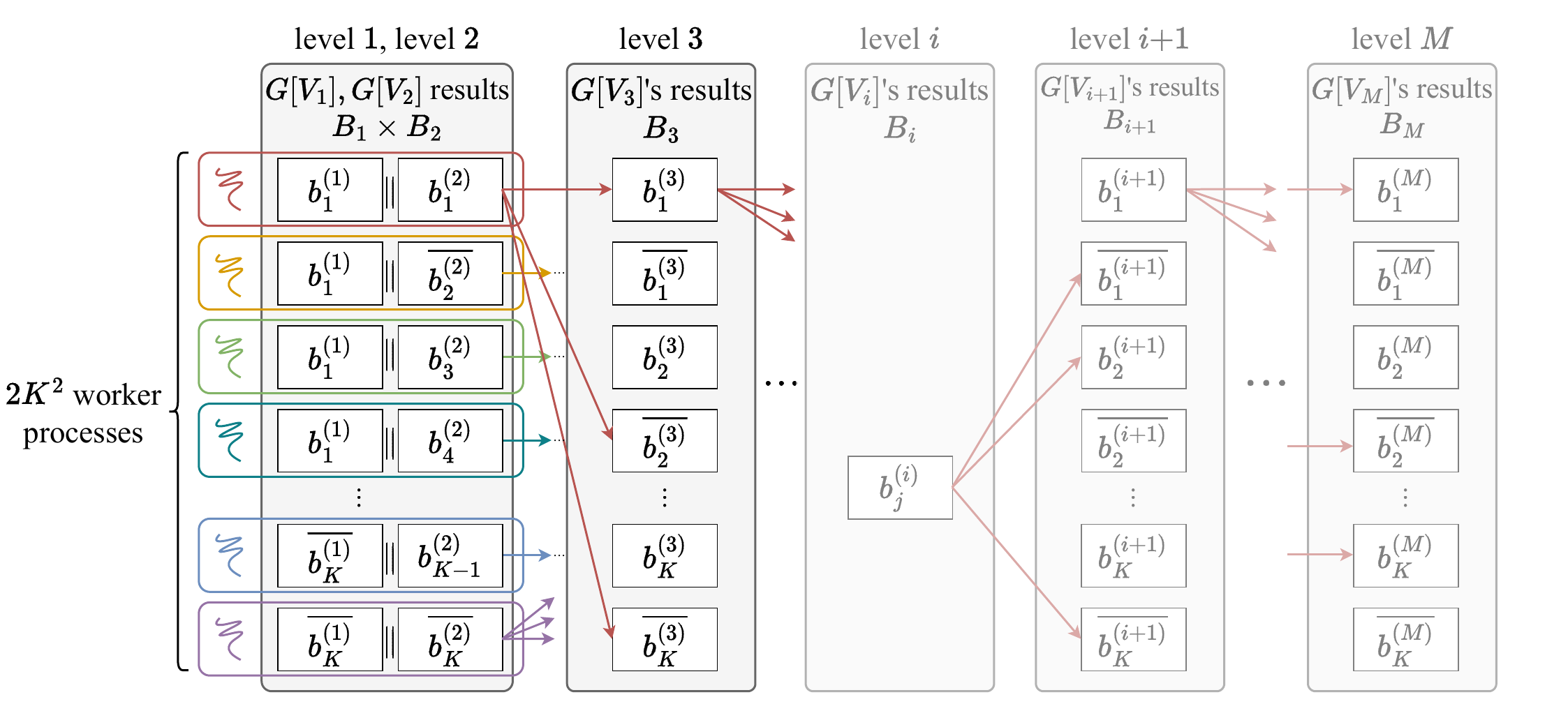}
    \caption{Level-aware parallel depth-first traversal merging. Starting at level $2$ increases parallelism compared to starting at level $1$ (\figurename~\ref{fig:level_aware_merge1}) by merging the results of $G[V_1]$ and $G[V_2]$ first. The right-hand side is dimmed to indicate that the subsequent process is identical to that shown in \figurename~\ref{fig:level_aware_merge1}.}
    \label{fig:level_aware_merge2}
\end{figure}

To efficiently explore the solution space formed by the Cartesian product of subgraph results, we employ a \emph{Level-Aware} merging strategy. This approach allows users to set the starting level of the merge phase, enabling finer-grained parallelism in searching the solution space.
\figurename~\ref{fig:level_aware_merge2} illustrates the level-aware merging process starting at level $2$, which increases parallelism compared to \figurename~\ref{fig:level_aware_merge1}, where the merge begins at the first level. This design enhances hardware utilization by expanding the number of worker processes from $2K$ to $2K^2$, thereby improving scalability when additional computational resources are available. In general, the number of worker processes is $2K^{\text{L}}$, where $\text{L}$ is the starting level of the merge phase. This allows for flexible parallel processing based on available hardware resources and user-defined parameters.

Algorithm~\ref{alg:level_aware_merge} details the proposed Level-Aware Parallel Merge methodology. It recursively constructs valid global configurations by traversing the solution space level by level, maintaining a current path that represents the partially constructed global solution. The algorithm employs parallel and level-aware merge techniques mentioned above, and finally returns the maximum cut value and the corresponding global configuration. The algorithm operates through several distinct phases.First, it initializes the maximum cut value and the corresponding configuration which track the best solution found during the traversal. The algorithm then generates starting points by merging the bitstring results of the first $L$ subgraphs, where $L$ is the user-defined starting level for the merge phase. This step creates a set of initial paths that will be expanded in parallel. Second, the algorithm spawns parallel processes to continue \emph{DepthFirstTraversal} from each starting path at level $L+1$ iteratively until it reaches the final level $M$. Each process explores the solution space by appending compatible bitstrings from the current subgraph's result set to the current path. When the travelsal reaches the final level $M$, the algorithm computes the cut and updates the maximum cut value if a better solution is found. Then, the algorithm waits for all processes to complete. Finally, the maximum cut value $C_{max}$ and the associated global configuration $V_{max}$ are returned.

\begin{algorithm}[htbp]
    \caption{Level-Aware Parallel Merge}\label{alg:level_aware_merge}
    \KwIn{$\{B_1, B_2, \dots, B_M\}$, list of subgraph results,\newline
        $L$, starting level of the merge phase for level-aware merging}
    \KwOut{$(V_{max}, C_{max})$, the maximum cut value and its associated cut configuration}
    \SetKwFunction{DepthFirstTraversal}{DepthFirstTraversal}
    \SetKwProg{Fn}{Function}{:}{}
    Initialize $V_{max} \gets 0$, $C_{max} \gets \text{null}$\;
    $\text{starting\_paths} \gets$ merge $B_1, \ldots, B_L$\Comment*[r]{Use L to merge bitstrings and create starting points}
    \BlankLine
    \tcc{Parallel depth-first traversal from level L}
    \ForEach{path in starting\_paths}{
        \textbf{Spawn Process:} \DepthFirstTraversal{L+1, path}\;
    }
    \BlankLine
    \Fn{\DepthFirstTraversal{level, current\_path}}{
        \If{level $= M$}{
            $V_{cut} \gets \textsc{CutVal}(\text{current\_path})$\;
            \If{$V_{cut} > V_{max}$}{
                $V_{max} \gets V_{cut}$; $C_{max} \gets \text{current\_path}$\;
            }
            \Return\;
        }
        \ForEach{compatible bitstring $b$ in $B_{level}$}{
            \DepthFirstTraversal{level $+ 1$, current\_path $\Vert$ $b$}\Comment*[r]{Bitstring concatenation}
        }
    }
    \BlankLine
    \textbf{Wait for all processes to complete}\;
    \Return{$(V_{max}, C_{max})$}\;
\end{algorithm}

\subsection{Performance Evaluation Metric}
\label{subsec:pei_framework}

\begin{figure}[htbp]
    \centering
    \includegraphics[width=0.5\textwidth]{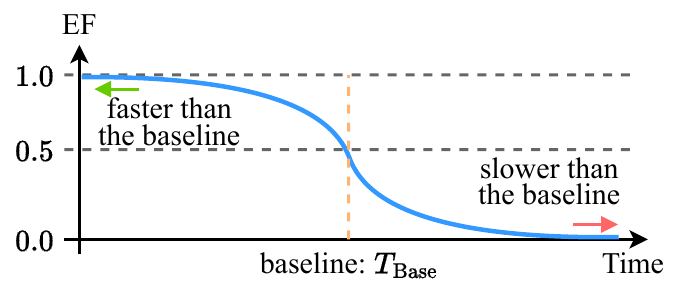}
    \caption{Visualization of the efficiency factor's value range.}
    \label{fig:efficiency_factor}
\end{figure}

To systematically evaluate the overall performance delivered by different approaches to the Max-Cut problem, the \emph{Performance Efficiency Index} is proposed as a unified metric to quantify the performance trade-off between solution quality and computational efficiency. 
Inspired by the \emph{Energy Delay Product}~\cite{LarosIII2013}, originally proposed by Horowitz~\cite{Horowitz} for evaluating trade-offs in digital circuit designs, PEI is tailored to the domain of quantum optimization, where gains in solution quality often come at the cost of increased execution time.
Formally, the PEI is defined as follows.
\[
    \text{PEI} = \text{AR} \times \text{EF} \times 100
    \label{eq:pei}
\]
This index integrates two key components: the \emph{Approximation Ratio (AR)}, which measures the quality of the Max-Cut solution relative to optimal or benchmark values, and the \emph{Efficiency Factor (EF)}, which quantifies computational efficiency with respect to baseline runtimes. By combining these dimensions, PEI offers a standardized and interpretable metric for assessing algorithm performance across various parameter settings and baseline methods, enabling multi-objective evaluation in a single score.

\[
    \text{Approximation Ratio: AR} = \frac{\text{CutVal}_\text{ALG}}{\text{CutVal}_\text{OPT}},
    \label{eq:approximation_ratio}
\]
where $\text{CutVal}_\text{ALG}$ denotes the cut value obtained by the evaluated algorithm, and $\text{CutVal}_\text{OPT}$ represents the optimal cut value of the original graph or the best-known value from existing methods.
The approximation ratio quantifies how close the obtained solution is to the optimal, with values ranging from 0 to 1, where $\text{AR} = 1$ indicates an optimal solution. In this work, for problems with 100 to 400 vertices, the best cut value is obtained from the GW method~\cite{gw}, which guarantees an approximation ratio of at least 0.878 in polynomial time. For larger instances, best-known solutions from state-of-the-art classical algorithms are used as reference baselines for evaluating solution quality.

\[
    \text{Efficiency Factor: EF } = \frac{1}{1+e^{\alpha\cdot(T_{\text{ALG}}-T_{\text{Base}})}},
    \label{eq:efficiency_factor}
\]
where $T_{\text{ALG}}$ is the execution time of the evaluated method, $T_{\text{Base}}$ is the baseline execution time, and $\alpha$ is a scaling parameter that controls the sensitivity of the efficiency ratio $e$ to timing differences. The efficiency ratio employs a sigmoid function that provides smooth transitions between performance regimes and bounded output values. As the illustration of \figurename~\ref{fig:efficiency_factor}, when $T_{\text{ALG}} = T_{\text{Base}}$, the efficiency ratio equals 0.5, representing performance parity. Values approaching 1.0 indicate significant computational acceleration (when $T_{\text{ALG}} \ll T_{\text{Base}}$), while values approaching 0.0 represent substantial performance degradation (when $T_{\text{ALG}} \gg T_{\text{Base}}$).

The sigmoid formulation offers several advantages over linear or logarithmic normalization: it naturally handles extreme timing variations without numerical instability, provides intuitive interpretation through its bounded range, and ensures balanced contribution to the overall PEI score regardless of the magnitude of timing differences. In this work, we set the scaling parameter to a small value, e.g., $\alpha = 0.001$, to ensure smooth transitions and prevent extreme timing variations from dominating the efficiency component.

\section{Evaluation}\label{sec:evaluation}

This section presents the experimental evaluation of the proposed ParaQAOA framework. The experimental setup and computational infrastructure are described in \sectionname~\ref{sec:experimental_setup}. We show that our framework is capable of managing the trade-off between solution quality and computational efficiency by configuring the framework's parameters in \sectionname~\ref{sec:effectiveness_of_parameter_configurations}. \sectionname~\ref{sec:performance_comparison} compares the performance of ParaQAOA with state-of-the-art QAOA-based Max-Cut solvers on small- and medium-scale graphs (fewer than 400 vertices), focusing on approximation ratio and execution time. Scalability results on large-scale instances (over 1,000 vertices) are presented in \sectionname~\ref{sec:scalability_results}. Finally, \sectionname~\ref{sec:pei_evaluation} evaluates performance efficiency using the proposed Performance Efficiency Index, which captures the trade-off between solution quality and computational cost.

\subsection{Experimental Setup}\label{sec:experimental_setup}

All experiments are conducted on a local high-performance computing system, with specifications provided in \tablename~\ref{tab:system_specs}. It is important to note that although the motherboard supports PCIe 5.0, the NVIDIA RTX 4090 GPUs operate at PCIe 4.0 speeds due to hardware limitations of the GPUs. Moreover, for a fair comparison, the classical Goemans-Williamson approximation algorithm (GW)~\cite{gw} (implemented by \cite{CQ}), Coupling QAOA (CQ)~\cite{CQ}\footnote{\url{https://github.com/LucidaLu/QAOA-with-fewer-qubits/tree/30a6a3fe24fe664281e17a3723573d1abf0b06df}.} and QAOA-in-QAOA (QAOA$^2$)~\cite{QiQ}~\footnote{\url{https://github.com/ZeddTheGoat/QAOA_in_QAQA/tree/7704cfd2c2cbfac58a11fbf4f8beebb1efb9c04c}.} are downloaded and evaluated on the same hardware platform.

\begin{table}[htbp]
    \centering
    \caption{Experimental computing infrastructure specifications.}
    \label{tab:system_specs}
    \begin{tabular}{ll}
        \toprule
        \textbf{Component} & \textbf{Specification}                                 \\
        \midrule
        CPU                & AMD Ryzen Threadripper 7960X (24-core)                 \\
        GPU                & 2$\times$ NVIDIA RTX 4090s each with 24 GB GDDR6X VRAM \\
        System Memory      & 256 GB (8$\times$ 32 GB DDR5 4800 MHz)                 \\
        Motherboard        & TRX50 AERO D with two PCIe 5.0 x16 slots               \\
        Operating System   & Ubuntu 20.04 LTS                                       \\
        Software           & CUDA 12.5, Python 3.12                                 \\
        Python Packages    & NetworkX 3.4.2, Numpy 2.0.2, Numba 0.60.0              \\
        \bottomrule
    \end{tabular}
\end{table}

We evaluate performance across various Erdős-Rényi random graph~\cite{erdds1959random} configurations, covering small-scale (20--26 vertices), medium-scale (100--400 vertices), and large-scale (1,000--16,000 vertices) instances, with edge probabilities set to 0.1, 0.3, 0.5, and 0.8. Ten test graphs are generated using the \texttt{gen\_erdos\_renyi\_graph} function from the NetworkX library~\cite{networkx}, each with a different random seed\footnote{Integer seeds from 0 to 9 are used for all graph generation configurations.} to ensure consistent and reproducible randomization across all experiments. For medium- and large-scale instances, a single fixed seed is used to manage the overall runtime of the experiments.
Due to inherent limitations, not all frameworks are applicable to all graph sizes. In the small-scale evaluation, we report results for GW, CQ, and QAOA$^2$. For medium-scale graphs (up to 400 vertices), only QAOA$^2$ and ParaQAOA are evaluated by using GW as the baseline. CQ is excluded. In the large-scale setting (up to 16,000 vertices), we compare QAOA$^2$ and ParaQAOA; however, QAOA$^2$ requires over 9 hours to solve graphs with 4,000 vertices, so its results for larger instances are extrapolated.

\subsection{Parameter Configurations for Managing the Quality-Efficiency Trade-off}\label{sec:effectiveness_of_parameter_configurations}

Our framework involves several key parameters: the number of QAOA solvers ($N_s$), the number of qubits per solver ($N$), the number of partitioned subgraphs ($M$), the number of QAOA solving rounds ($T$), the candidate solution preservation parameter ($K$), and the starting level ($L$) in the merging process. These parameters jointly influence the trade-off between solution quality and computational efficiency. This section evaluates their impact on solution quality and execution time.
We categorize the parameters into three types: hardware-dependent, input-dependent, and tunable. We describe the configuration of these parameters and present the experimental results showing how the tunable parameter influences the performance of solving the Max-Cut problem.
To assess the effectiveness of tunable parameters, we analyze their impact on the obtained cut value and execution time for graphs with 200 and 600 vertices across various edge probabilities.

\paragraph{Hardware-Dependent Parameters}
The hardware-dependent parameters include $N_s$ and $N$, which are determined by the available computational resources. Based on the hardware specifications (\tablename~\ref{tab:system_specs}), we deploy up to 12 concurrent QAOA solver instances per GPU, yielding $N_s = 24$ solvers in total. This configuration maximizes physical CPU core utilization without overloading the system. Each solver is allocated up to $N = 26$ qubits, allowing it to process subgraphs with up to 26 vertices, while consuming less than 1~GB of GPU memory. The remaining memory is reserved for intermediate data during execution.

\paragraph{Input-Dependent Parameters}
The input-dependent parameters include the number of partitioned subgraphs ($M$) and the number of QAOA solving rounds ($T$). The value of $M$ is determined by the input graph size $|V|$ and the number of qubits $N$ allocated per solver. Specifically, $M$ is set to $|V| / (N - 1)$, where the subtraction accounts for one shared node between adjacent subgraphs. The number of solving rounds $T$ is set to $M / N_s$, ensuring that each QAOA solver handles a balanced workload and that all solvers are fully utilized throughout the execution.

\paragraph{Tunable Parameters}
The tunable parameters, $K$ and $L$, allow further control over the trade-off between solution quality and computational efficiency. Parameter $K$ specifies the number of top-$K$ high-probability bitstrings preserved during the parallel QAOA execution stage, influencing the diversity of candidate solutions. Parameter $L$ defines the starting level in the merging process, affecting how early subgraph solutions begin to combine. These parameters can be adjusted to align the framework's behavior with application-specific requirements and hardware capabilities.

\begin{figure}[htbp]
    \centering
    \includegraphics[width=0.9\textwidth]{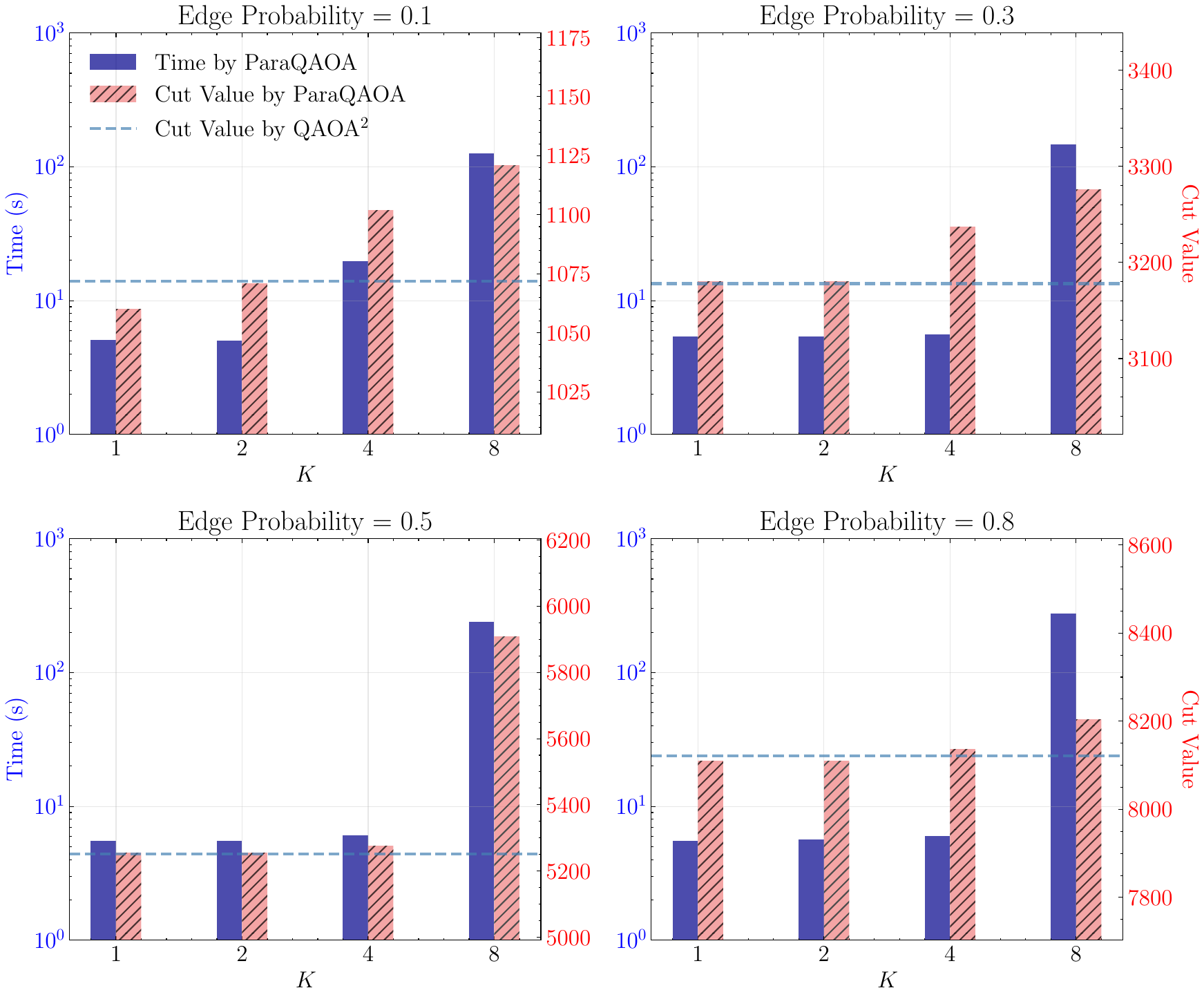}
    \caption{Cut values and execution times of ParaQAOA for varying filtering parameter $K$ on Erdős-Rényi graphs with 200 vertices across different edge probabilities.}
    \label{fig:different_K}
\end{figure}

Parameter $K$ serves as a tunable trade-off control in our framework. It allows users to balance solution quality and computational efficiency according to application-specific needs. To evaluate this trade-off, we analyze the impact of $K$ on the cut value and execution time. \figurename~\ref{fig:different_K} presents the results for varying $K$ values, which determine the number of top-$K$ high-probability bitstrings retained during the parallel QAOA execution.
The x-axis represents $K$, while the left y-axis shows execution time and the right y-axis indicates the achieved cut values by ParaQAOA. The results indicate that increasing $K$ improves solution quality but incurs higher execution time, reflecting a trade-off between accuracy and efficiency. In most cases, ParaQAOA achieves cut values comparable to QAOA$^2$ with $K=1$ or $K=2$, while maintaining substantially lower execution times, demonstrating its ability to balance quality and efficiency through parameter tuning.

\begin{figure}[htbp]
    \centering
    \includegraphics[width=0.9\textwidth]{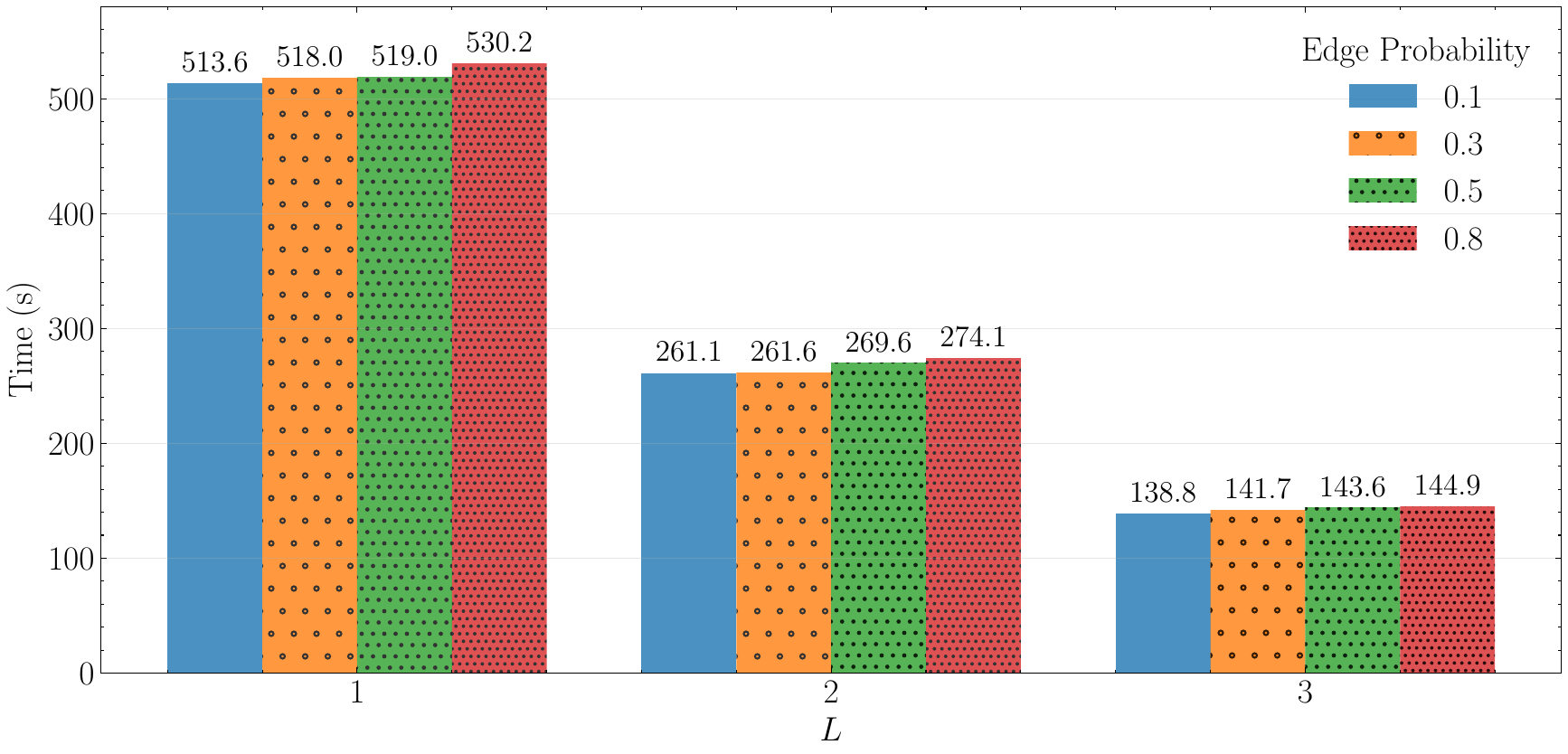}
    \caption{Execution time of ParaQAOA with varying starting point $L$ on Erdős-Rényi graphs with 600 vertices across different edge probabilities with $K=2$.}
    \label{fig:different_L}
\end{figure}

The parameter $L$ controls the starting level in the merging process and directly affects parallelism during post-processing. An appropriate choice of $L$ improves computational efficiency by aligning the parallel merge operations with available CPU cores. A recommended configuration is to set $2K^L$ close to the number of physical CPU cores, which ensures efficient use of hardware without incurring system overhead. \figurename~\ref{fig:different_L} illustrates the performance of ParaQAOA with $L$ ranging from 1 to 3. The parameter $L$ determines the number of processes executed in parallel. For example, when $L=1$, four processes are spawned; when $L=2$, eight processes; and when $L=3$, sixteen processes. The results show that doubling the number of processes reduces the runtime by approximately half. This highlights the impact of $L$ on execution efficiency and underscores the framework's scalability potential on modern multi-core and many-core systems~\cite{JEFFERS20163}.

\subsection{Performance Evaluation on Small- and Medium-scale Graphs}\label{sec:performance_comparison}

We begin our evaluation with small-scale Max-Cut instances, specifically Erdős-Rényi random graphs with 20 to 30 vertices. The results are summarized in \tablename~\ref{tab:small_scale}, which compares the approximation ratios and execution times of ParaQAOA against CQ, and QAOA$^2$.
The results show that ParaQAOA achieves competitive approximation ratios while significantly outperforming the other methods in terms of execution time. For instance, on a 20-vertex graph with edge probability 0.1, ParaQAOA achieves an approximation ratio of 82.5\% in just 0.85 seconds, while QAOA$^2$ takes 2.46 seconds and CQ takes 106.87 seconds to achieve similar results. As the graph size and edge probability increases, ParaQAOA continues to demonstrate superior performance, maintaining high approximation ratios while keeping execution times low.
The results demonstrate that ParaQAOA achieves competitive approximation ratios while significantly outperforming the other methods in terms of execution time.

\begin{table}[htbp]
    \centering
    \caption{Comparison of execution runtime and approximation ratio on small-scale Max-Cut instances with varying graph sizes ($|V|$) and edge probabilities ($P$). AR values are calculated by dividing the obtained cut value by the optimal cut value, determined by a brute-force method.}
    \label{tab:small_scale}
    \begin{tabular}{ccccc|ccc}
        \toprule
        \multirow{2}{*}{\textbf{$P$}} & \multirow{2}{*}{\textbf{$|V|$}} & \multicolumn{3}{c|}{\textbf{Runtime (s)}} & \multicolumn{3}{c}{\textbf{AR (\%)}}                                         \\
                                      &                                 & QAOA$^2$                                  & CQ                                   & ParaQAOA & QAOA$^2$ & CQ   & ParaQAOA \\
        \midrule
        \multirow{4}{*}{0.1}
                                      & 20                              & \p{0}2.46                                 & \p{0}106.87                          & 0.85     & 84.6     & 94.8 & 82.5     \\
                                      & 22                              & \p{0}2.56                                 & \p{0}181.51                          & 0.86     & 87.0     & 95.6 & 85.9     \\
                                      & 24                              & \p{0}2.93                                 & \p{0}280.82                          & 0.87     & 83.7     & 96.1 & 90.7     \\
                                      & 26                              & \p{0}2.92                                 & 1044.23                              & 0.89     & 84.7     & 96.8 & 81.4     \\
        \midrule
        \multirow{4}{*}{0.3}
                                      & 20                              & \p{0}8.18                                 & \p{0}151.88                          & 0.85     & 93.8     & 97.7 & 89.4     \\
                                      & 22                              & \p{0}8.89                                 & \p{0}246.90                          & 0.85     & 92.1     & 97.9 & 89.5     \\
                                      & 24                              & \p{0}8.71                                 & \p{0}488.65                          & 0.87     & 91.5     & 98.4 & 90.9     \\
                                      & 26                              & \p{0}9.55                                 & 1531.19                              & 0.91     & 90.9     & 98.4 & 86.9     \\
        \midrule
        \multirow{4}{*}{0.5}
                                      & 20                              & 16.18                                     & \p{0}179.31                          & 0.86     & 95.6     & 98.4 & 92.0     \\
                                      & 22                              & 17.16                                     & \p{0}290.58                          & 0.88     & 95.1     & 98.6 & 93.3     \\
                                      & 24                              & 18.92                                     & \p{0}501.91                          & 0.90     & 94.5     & 98.7 & 93.9     \\
                                      & 26                              & 19.38                                     & 1590.81                              & 0.91     & 94.1     & 98.8 & 93.0     \\
        \midrule
        \multirow{4}{*}{0.8}
                                      & 20                              & 33.29                                     & \p{0}219.46                          & 0.87     & 96.9     & 98.5 & 94.6     \\
                                      & 22                              & 35.39                                     & \p{0}367.41                          & 0.89     & 95.5     & 99.1 & 96.1     \\
                                      & 24                              & 34.23                                     & \p{0}586.28                          & 0.90     & 96.4     & 99.1 & 97.2     \\
                                      & 26                              & 37.49                                     & 1592.60                              & 0.91     & 95.4     & 99.1 & 95.6     \\
        \bottomrule
    \end{tabular}
\end{table}

\begin{figure}[htbp]
    \centering
    \includegraphics[width=0.8\textwidth]{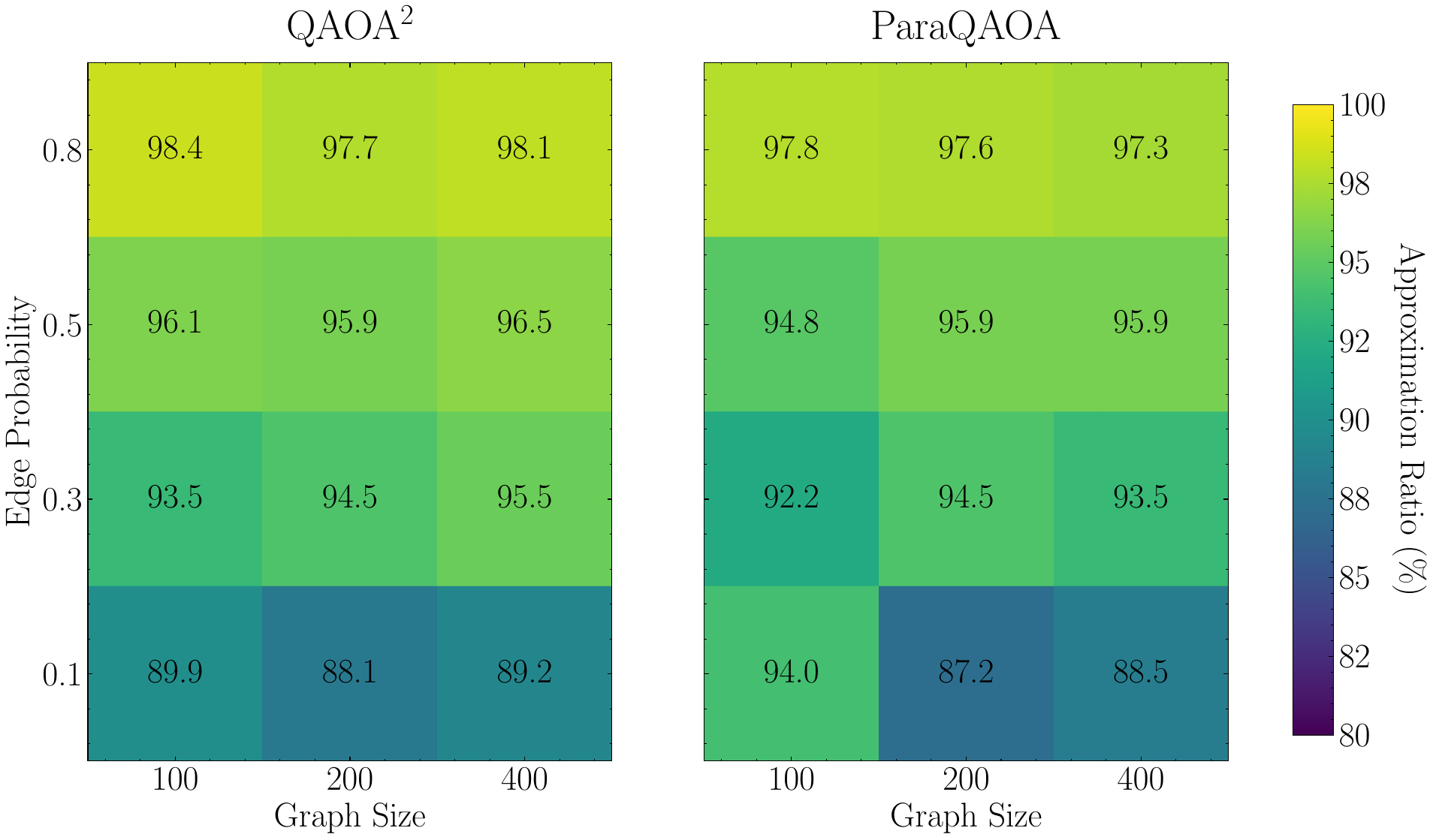}
    \caption{Approximation ratio, computed using cut values from GW, shown as a heatmap comparing QAOA$^2$ and ParaQAOA.}
    \label{fig:ours_qiq_ar_heatmap}
\end{figure}

For medium-scale problems, we compare the performance of our ParaQAOA framework with QAOA$^2$, a state-of-the-art QAOA-based Max-Cut solver. CQ is excluded from this evaluation due to its long runtime (exceeding 8 hours for a 30-vertex graph), its restriction to bipartite graphs, and its inability to handle larger instances.
The GW algorithm serves as the baseline, as brute-force methods are infeasible at this scale. \tablename~\ref{tab:runtime_comparison} and \figurename~\ref{fig:ours_qiq_ar_heatmap} summarize the performance results delivered by ParaQAOA and QAOA$^2$.

\tablename~\ref{tab:runtime_comparison} reports execution time results and demonstrates the computational efficiency of ParaQAOA. Our framework consistently outperforms QAOA$^2$ across all configurations, with speedups increasing with graph edge probability. For example, on a 100-vertex graph with edge probability 0.1, ParaQAOA achieves a $112.1\times$ speedup, which scales to $1652.2\times$ on a 400-vertex graph with edge probability 0.8.
This performance trend underscores key algorithmic differences. QAOA$^2$ exhibits exponential growth in computation time with increasing graph edge probability due to its exhaustive sub-solution enumeration. In contrast, ParaQAOA's runtime is dominated by the performance of individual QAOA solvers, making it significantly less sensitive to graph complexity.

\begin{table}[htbp]
    \centering
    \caption{Execution time comparison of QAOA$^2$ and ParaQAOA on medium-scale Max-Cut instances with varying graph sizes ($|V|$) and edge probabilities ($P$).}
    \label{tab:runtime_comparison}
    \begin{tabular}{ccccc}
        \toprule
        \textbf{\makecell{$P$}} & \textbf{\makecell{$|V|$}} & \textbf{\makecell{QAOA$^2$                                             \\Runtime (s)}} & \textbf{\makecell{ParaQAOA\\Runtime (s)}} & \textbf{Speedup} \\
        \midrule
        \multirow{3}{*}{0.1}
                                & 100                       & \phantom{00}668.0          & \phantom{0}6.0 & \phantom{0}112.1$\times$ \\
                                & 200                       & \phantom{0}1128.6          & \phantom{0}7.3 & \phantom{0}155.2$\times$ \\
                                & 400                       & \phantom{0}2158.8          & \phantom{0}8.7 & \phantom{0}247.0$\times$ \\
        \midrule
        \multirow{3}{*}{0.3}
                                & 100                       & \phantom{00}888.8          & \phantom{0}6.4 & \phantom{0}138.0$\times$ \\
                                & 200                       & \phantom{0}1754.1          & \phantom{0}7.7 & \phantom{0}227.2$\times$ \\
                                & 400                       & \phantom{0}3320.7          & \phantom{0}9.9 & \phantom{0}334.4$\times$ \\
        \midrule
        \multirow{3}{*}{0.5}
                                & 100                       & \phantom{0}1753.5          & \phantom{0}6.5 & \phantom{0}268.1$\times$ \\
                                & 200                       & \phantom{0}2937.7          & \phantom{0}7.9 & \phantom{0}372.8$\times$ \\
                                & 400                       & \phantom{0}6943.8          & 10.2           & \phantom{0}679.4$\times$ \\
        \midrule
        \multirow{3}{*}{0.8}
                                & 100                       & \phantom{0}4659.4          & \phantom{0}6.6 & \phantom{0}706.0$\times$ \\
                                & 200                       & \phantom{0}8591.3          & \phantom{0}8.2 & 1051.6$\times$           \\
                                & 400                       & 17001.0                    & 10.3           & 1652.2$\times$           \\
        \bottomrule
    \end{tabular}
\end{table}

\figurename~\ref{fig:ours_qiq_ar_heatmap} presents the approximation ratio comparison between ParaQAOA and QAOA$^2$. Two key observations are drawn from the results.
First, both methods yield lower ARs on graphs with low edge probability due to the increased influence of individual edges on the cut value. Since both frameworks employ random partitioning and disregard inter-subgraph edges, sparse graphs are more susceptible to approximation degradation. In contrast, on denser graphs, where individual edges contribute less significantly, both methods achieve ARs approaching that of the GW algorithm.
Second, the worst-case AR degradation of our framework relative to QAOA$^2$ is within 2\%, while the typical difference remains around 1\%. In several configurations, ParaQAOA even surpasses QAOA$^2$ in solution quality.

\subsection{Scalability Results on Large-scale Graphs}\label{sec:scalability_results}

\begin{figure}[htbp]
    \centering
    \begin{subfigure}[b]{0.49\textwidth}
        \centering
        \includegraphics[width=\textwidth]{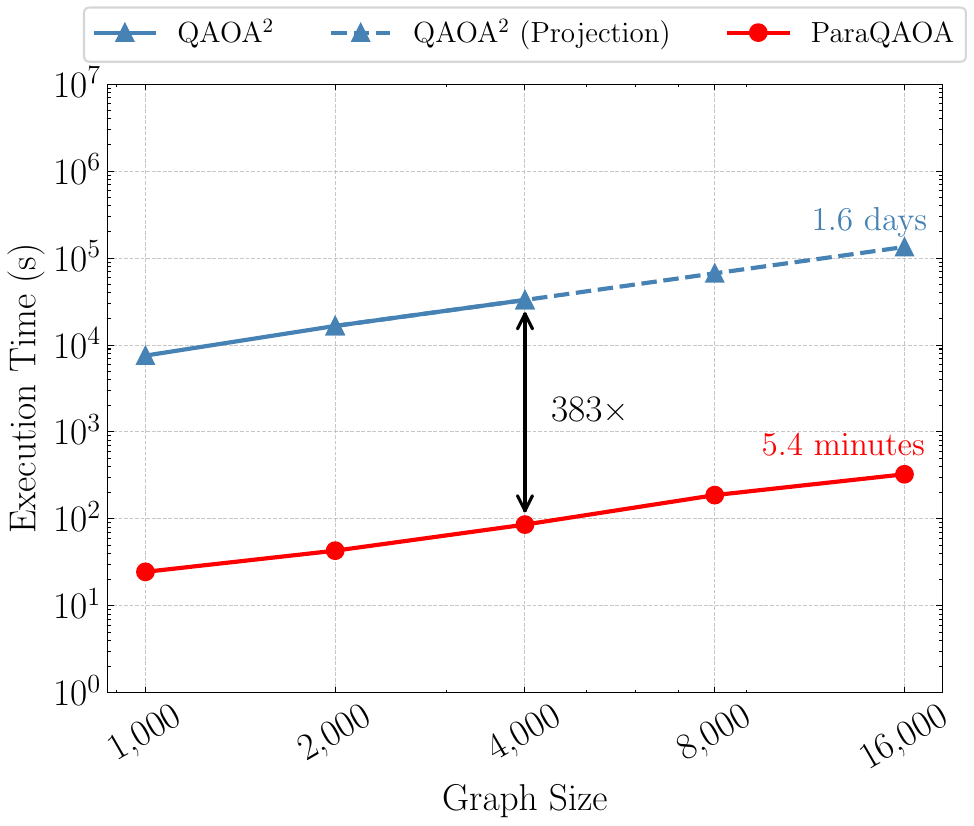}
        \caption{Edge Probability = 0.1}
        \label{fig:execution_time_comparison_01}
    \end{subfigure}
    \,
    \begin{subfigure}[b]{0.49\textwidth}
        \centering
        \includegraphics[width=\textwidth]{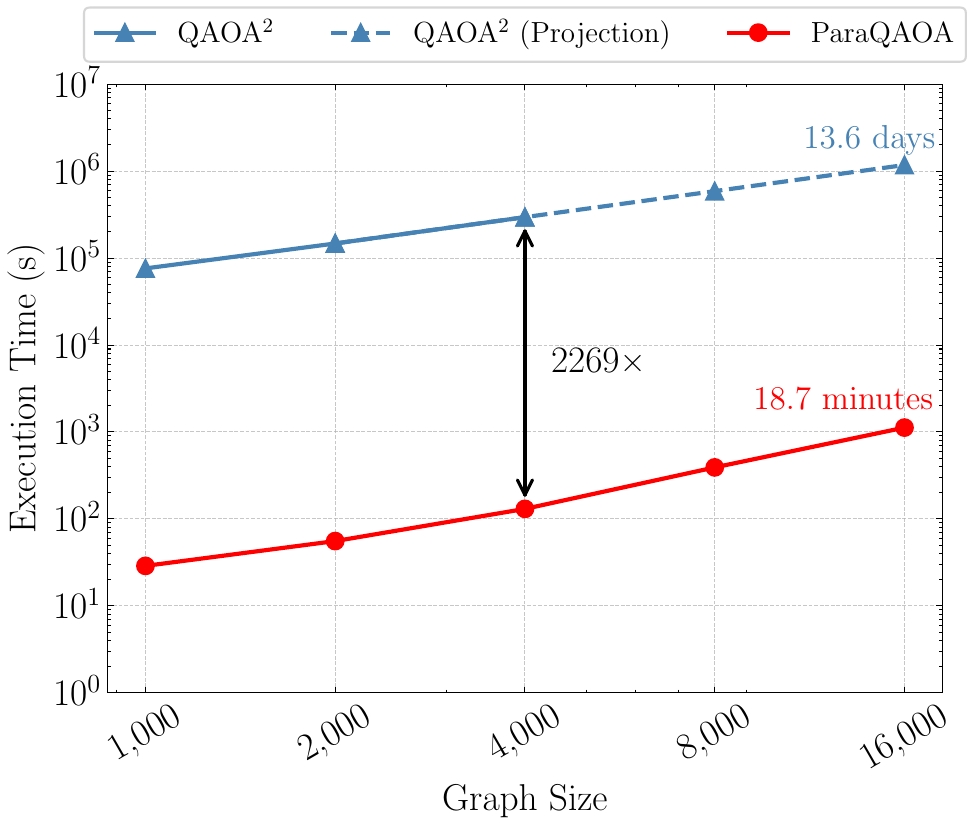}
        \caption{Edge Probability = 0.8}
        \label{fig:execution_time_comparison_08}
    \end{subfigure}
    \caption{Scalability analysis comparing execution times between QAOA$^2$ and ParaQAOA for large-scale Max-Cut problems.}
    \label{fig:execution_time_comparison_full}
\end{figure}

To evaluate scalability, we extended our experiments to large-scale graph instances ranging from 1,000 to 16,000 vertices. At this scale, only QAOA$^2$ and our ParaQAOA framework are capable of solving the problems. \figurename~\ref{fig:execution_time_comparison_full} shows execution time trends for representative low (0.1) and high (0.8) edge probability configurations; intermediate edge probability cases exhibit similar scaling patterns.
Due to the high computational cost of QAOA$^2$ at large scales, its execution time was measured only for graphs with up to 4,000 vertices. For larger instances, we applied linear regression to extrapolate execution time (denoted by ``Projecttion'') based on the observed relationship between the number of vertices and QAOA$^2$'s runtime.

Our analysis yields two key observations. First, as problem size increases, the performance of QAOA$^2$ is significantly affected by edge density. Specifically, increasing the edge probability from 0.1 to 0.8 results in approximately a 10-fold increase in execution time for graphs with the same number of vertices. In contrast, our framework demonstrates much greater robustness to edge density, with execution time increasing by at most 1.5$\times$ across the same range of edge probabilities for graphs with 1,000 to 4,000 vertices.
Second, our framework consistently outperforms QAOA$^2$ in terms of computational efficiency and achieves speedups ranging from 300$\times$ to 2,000$\times$ as problem complexity increases.
In addition to relative performance, we present absolute execution times to emphasize the practical applicability of our framework. For graphs with 16,000 vertices, our method completes in minutes, whereas QAOA$^2$ requires several days, rendering it impractical for real-world use at this scale.

\subsection{Performance Efficiency Index Evaluation}\label{sec:pei_evaluation}

\begin{figure}[htbp]
    \centering
    \includegraphics[width=\textwidth]{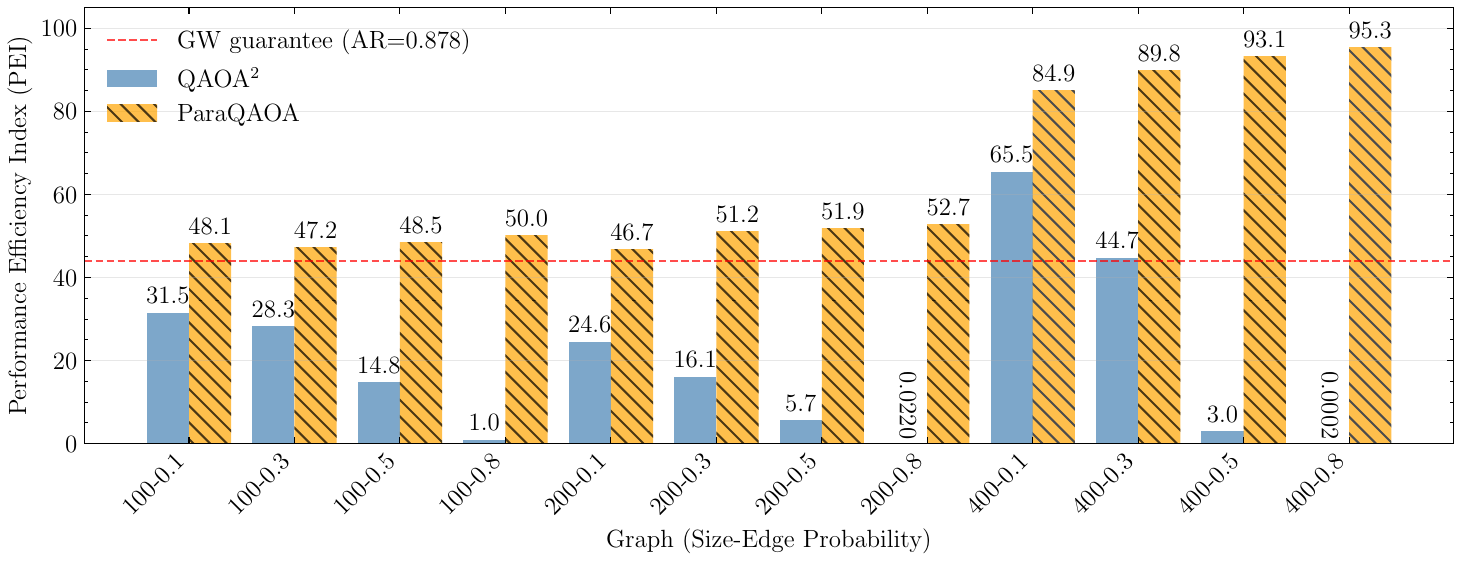}
    \caption{Comparison of the Performance Efficiency Index between QAOA$^2$ and ParaQAOA across different graph sizes and edge probabilities, using the GW algorithm as the baseline for cut values and runtimes.}
    \label{fig:plot_PEI}
\end{figure}

As presented in \sectionname~\ref{subsec:pei_framework}, PEI is a novel metric that combines approximation ratio and execution time into a single index to evaluate the performance of different solutions to the Max-Cut problem.
\figurename~\ref{fig:plot_PEI} plots PEI values across various medium-scale graph configurations, covering vertex counts of 100, 200, and 400, and edge probabilities ranging from 0.1 to 0.8. The GW algorithm serves as the baseline for computing AR and EF in the proposed PEI metric, with $\alpha=0.001$ used to ensure smooth scaling of runtime data.
Our proposed framework consistently outperforms QAOA$^2$ in all tested configurations. Notably, the advantage becomes more pronounced as graph complexity increases, either through larger vertex counts or higher edge densities. In addition, our method surpasses the approximation performance guarantee of the GW algorithm (represented by the horizontal red dashed line), highlighting its ability to achieve a better trade-off between solution quality and computational efficiency.

\begin{figure}[htbp]
    \centering
    \includegraphics[width=0.6\textwidth]{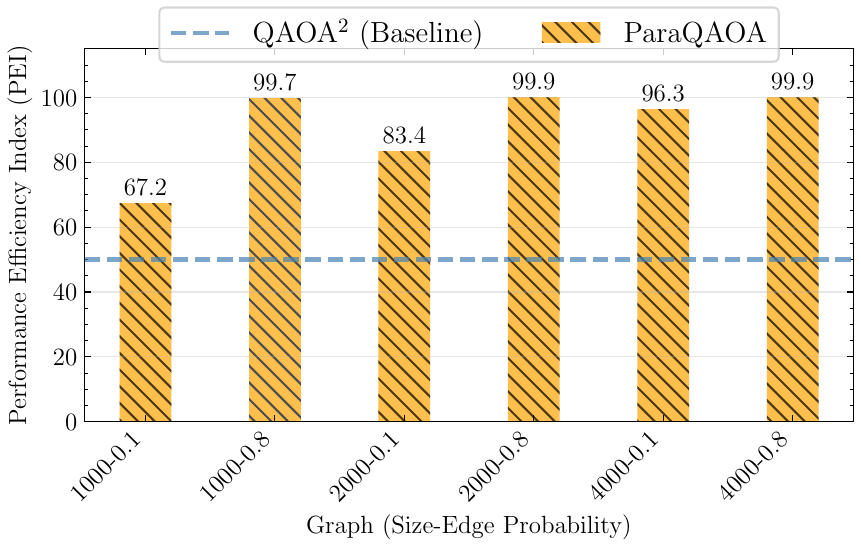}
    \caption{Performance Efficiency Index of ParaQAOA across different graph sizes and edge probabilities, using QAOA$^2$ as the baseline for cut values and runtimes.}
    \label{fig:plot_PEI_large}
\end{figure}

Additionally, we extend the PEI evaluation to large-scale Max-Cut instances, as shown in \figurename~\ref{fig:plot_PEI_large}. Due to the size of these graphs and the limitation of the existing implementation, the GW algorithm is no longer applicable. Instead, we use QAOA$^2$ as the baseline, treating its best cut values as reference. In addition, the parameter $\alpha$ is set to $0.0001$ to ensure smooth scaling of runtime data and to reflect the runtime gap observed at this problem scale. The results show that ParaQAOA continues to deliver strong performance efficiency across all tested large-scale graphs.

\section{Conclusion}\label{sec:conclusion}

This paper presents the ParaQAOA framework, a novel hybrid quantum-classical approach that successfully addresses the fundamental trade-off between solution quality and computational efficiency in large-scale combinatorial optimization. Through systematic algorithmic innovations and comprehensive experimental validation, we demonstrate that practical quantum-inspired optimization can achieve both scalability and performance suitable for real-world deployment.
The introduction of the Performance Efficiency Index offers a generalizable framework for evaluating trade-offs between solution quality and computational efficiency across optimization algorithms. 

Despite these strengths, several limitations suggest directions for future research. Current randomized partitioning may underperform on structured graphs, motivating exploration of adaptive partitioning techniques. While the method is evaluated on Max-Cut, extending it to other QUBO problems (e.g., TSP, graph coloring) could demonstrate broader applicability. Hardware-specific optimization and noise-aware circuit designs are also promising directions.


\begin{acks}
This work was supported in part by National Science and Technology Council, Taiwan, under Grants 113-2119-M-002 -024 and 114-2221-E-006 -165 -MY3. We thank to National Center for High performance Computing (NCHC), High Performance and Scientific
Computing Center at National Taiwan University, and Inventec for providing computational and storage resources. We thank the financial supports from the Featured Area Research Center Program within the framework of the Higher Education Sprout Project by the Ministry of Education (114L900903).
\end{acks}

\bibliographystyle{ACM-Reference-Format}
\bibliography{paper}

\end{document}